\documentclass{article}

\usepackage{arxiv}

\usepackage[utf8]{inputenc} 
\usepackage[T1]{fontenc}    
\usepackage{hyperref}       
\usepackage{url}            
\usepackage{booktabs}       
\usepackage{amsfonts}       
\usepackage{nicefrac}       
\usepackage{microtype}      
\usepackage{lipsum}		
\usepackage{graphicx}
\usepackage{natbib}
\usepackage{doi}

\usepackage{multirow}%
\usepackage{amsmath,amssymb}%
\usepackage{amsthm}%
\usepackage{mathrsfs}%
\usepackage[title]{appendix}%
\usepackage{xcolor}%
\usepackage{textcomp}%
\usepackage{manyfoot}%
\usepackage{booktabs}%
\usepackage{algorithm}%
\usepackage{algorithmicx}%
\usepackage{algpseudocode}%
\usepackage{listings}%
\usepackage{subcaption}%

\title{Transition from laminar to turbulent pipe flow as a process of growing material instabilities}

\author{ \href{https://orcid.org/0000-0002-4932-5960}{\includegraphics[scale=0.06]{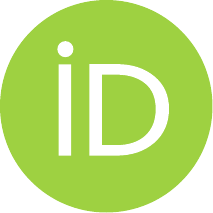}\hspace{1mm}Saptarshi Kumar Lahiri}\thanks{ Alternate mail - \texttt{lahiriayan22@gmail.com}---\emph{} Funded by Israel Science Foundation} \\
	Faculty of Civil and Environmental Engineering\\
	Technion Israel Institute of Technology\\
	Haifa, IL - 3200003 \\
	\texttt{saptarshi@campus.technion.ac.il} \\
	\And
	\href{ https://orcid.org/0000-0002-2770-3916}{\includegraphics[scale=0.06]{orcid.pdf}\hspace{1mm}Konstantin Volokh} \\
	Faculty of Civil and Environmental Engineering\\
	Technion Israel institute of Technology\\
	Haifa, IL - 3200003 \\
	\texttt{cvolokh@technion.ac.il} \\
}

\renewcommand{\shorttitle}{\textit{arXiv} Template}

\hypersetup{
pdftitle={A template for the arxiv style},
pdfsubject={q-bio.NC, q-bio.QM},
pdfauthor={David S.~Hippocampus, Elias D.~Striatum},
pdfkeywords={First keyword, Second keyword, More},
}

\begin{document}
\maketitle

\begin{abstract}
In this work, we simulate the transition to turbulence in the pipe flow based on the modified NS theory incorporating the viscous fluid strength in the constitutive equations. The latter concept enriches theory by allowing for material instabilities in addition to the kinematic ones. We present results of comparative numerical simulations based on the classical NS model and the NS model enhanced with the finite viscous strength. As expected, simulations based on the classical NS model exhibit stable laminar flow in contrast to experimental observations. Conversely, simulations based on the modified NS model with viscous strength exhibit instabilities and transition to turbulence per experimental observations. The transition to turbulence is triggered by the growing material instabilities.
\end{abstract}

\section{\label{sec:1}Introduction}
Considering pipe flow, Osborne Reynolds \cite{reynolds1883xxix, reynolds1895iv} introduced a non-dimensional parameter $Re = \rho v D /\eta$, in which $\rho$ is the mass density of the fluid, $v$ is the mean velocity, $D$ is the diameter of the pipe, and ${\eta}$ is the viscosity of the fluid. Reynolds assumed that the transition from laminar to turbulent flow occurred at some critical number $Re_{\text{cr}}$. Experimental estimates of this critical number for water vary from 1700 and 2300 \cite{faisst2004sensitive, eckhardt2007turbulence} to 3000 \cite{eckhardt2009introduction}. Recently, Avila et al. \cite{avila2011onset} experimentally studied laminar flow through a 15 m long pipe with a diameter of 4 mm and observed that at $Re_{\text{cr}}\approx 2400$ the laminar flow became unstable and indicated transition to turbulence.

Theoretical explanations of the transition to turbulence have a long history \cite{tanner1998rheology, manneville2015transition, manneville2016transition, barkley2016theoretical, eckert2022turbulenceproblem, eckert2022turbulence-anodyssey, feigenbaum1978quantitative, gollub1975onset} and debates proceed nowadays. Unfortunately, the linear instability analysis cannot capture the onset of failure of the laminar flow, which is observed in experiments. The linear instability analysis tracks the evolution of infinitesimal or small perturbations, e.g., molecular fluctuations, which always exist. Such perturbations do not develop into transitional flow according to the Navier-Stokes theory \cite{romanov1973stability, landau2013fluid}.

To accommodate the experimentally observed transition to turbulence within the framework of the Navier-Stokes theory, some authors assume the existence of finite or large perturbations in the laminar flow \cite{peixinho2006decay, eckhardt2007turbulence, eckhardt2009introduction}. The physical grounds for such an assumption are disputable. Interestingly, Wygnanski and Champagne \cite{wygnanski1973transition} observed that the enforcement of the finite perturbations did not always ensure instability.

We note in passing that the conventional NS approach also struggles to explain the physical mechanism of the drag reduction when a small amount of polymer molecules is added to a Newtonian solvent as reported widely in the literature \cite{toms1949some, boelens2016rotational, bewersdorff1995drag, nadolink1995bibliography, ptasinski2003turbulent, xi2016marginal, lumley1969drag}.

The difficulty of the classical NS theory to explain some of the experimentally observed transitions to turbulence led Volokh \cite{volokh2009investigation, volokh2013navier, volokh2018explanation} to assume that considerations of purely kinematic instabilities of the laminar flow were not enough and material instabilities of the flow should be taken into account as well. The latter extension of the NS theory can be done, for example, by the enforcement of finite viscous strength in the constitutive law. Physically, it means that viscosity drops at the critical strain rate because internal friction breaks down leading to instability of the laminar flow. The viscous stress corresponding to the critical strain rate is termed the viscous strength of the fluid. It is a material property.

Quite amazingly, the recent molecular dynamic simulations by Raghavan and Ostoja-Starzewski \cite{raghavan2017shear-thinning} confirm the assumption of viscous strength in the continuum mechanics formulation.

Despite its theoretical promise and some analytical results \cite{volokh2009investigation, volokh2013navier, volokh2018explanation}, the proposed modified NS constitutive model with viscous strength was not examined in specific numerical simulations of a three-dimensional pipe flow. This gap is filled in the present work. Results of numerical simulations are reported below, which show the transition to turbulence in the pipe flow. This transition is due to the process of growing material instabilities.

\section{\label{sec:2}Navier-Stokes model enhanced with viscous strength}

In this section, we summarize the governing equations. They include the linear momentum balance in the standard form
\begin{equation}
\rho\frac{\partial \mathbf{v}}{\partial t} + \rho(\text{grad}\mathbf{v})\cdot\mathbf{v} = -\text{grad}p +\text{div}\boldsymbol{\tau},
\label{eq_momt}
\end{equation}
where $\rho$ is the mass density; $\mathbf{v}$ is the velocity obeying the incompressibility condition $\text{div}\mathbf{v}=0$; $t$ is time; $p$ is the hydrostatic pressure; and $\boldsymbol{\tau}$ is the viscous stress tensor.

The governing equations also include the constitutive law. For Newtonian fluids, the viscous stress is proportional to the strain rate
\begin{equation}
\boldsymbol{\tau} = 2\eta \mathbf{D},
\label{eq_tau}
\end{equation}
where $\eta$ is the fluid viscosity and $\mathbf{D} = \left(\text{grad}\mathbf{v} + \text{grad}\mathbf{v}^{\text{T}}\right)/2$ is the strain rate tensor equal to the symmetric part of the velocity gradient. 

In the viscous strength model \cite{volokh2009investigation}, a finite strain rate limiter is incorporated in the viscosity function that remains constant until a critical point, beyond which the viscosity drops to zero representing a breakdown of contacts between fluid layers
\begin{equation}
\eta^* = \eta\text{exp} \left[ - \left(\frac{\mathbf{D}:\mathbf{D}}{\phi^2}\right)^m\right],
\label{eq_eta_vsm}
\end{equation}
where $\mathbf{D}:\mathbf{D} = \text{tr}\left[\mathbf{D}\mathbf{D}^{\text{T}}\right]$ is the squared equivalent scalar strain rate; $m$ is a constant; and $\phi$ is the critical equivalent strain rate, which is a saturation limit for the material stability of the fluid.

The viscosity function has essentially two modes -- nonzero constant, which describes a classical Newtonian fluid, and zero constant, corresponding to the ideal fluid with negligible internal friction:
\begin{equation}
\eta^*=
\begin{cases}
\eta	& ~ \text{when} ~ \mathbf{D}:\mathbf{D} < \phi^2 \\
0		& ~ \text{when} ~ \mathbf{D}:\mathbf{D} > \phi^2,
\end{cases}
\label{eq_vsm_eta1}
\end{equation}

\begin{figure}[!h]
\centering
	\begin{subfigure}{0.5\textwidth}
	\includegraphics[trim={0.2cm 0.2cm 0.15cm 1cm}, clip, width=\textwidth]{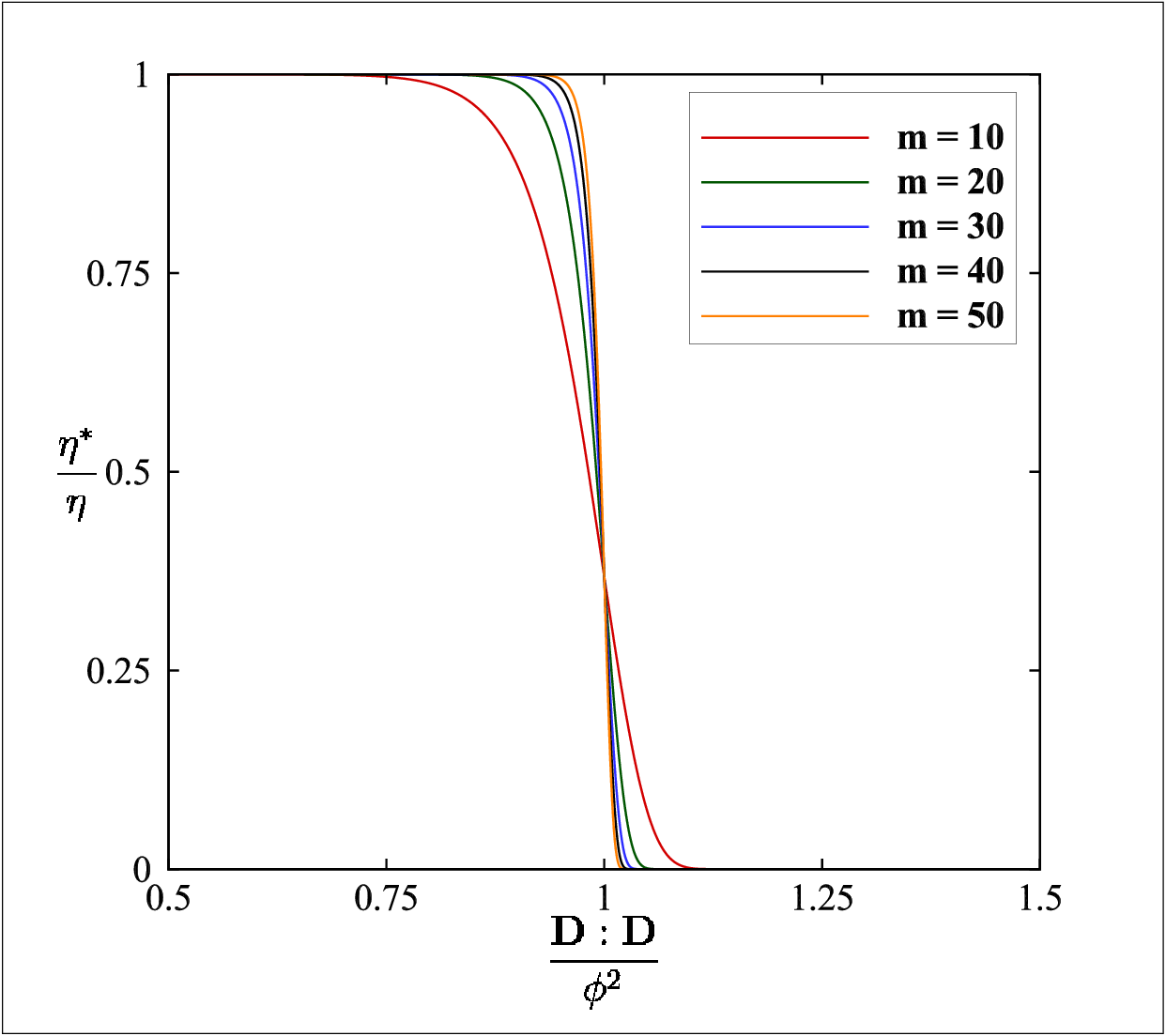}\vspace*{-0.1cm}
	\caption{\label{Fig_eta} viscosity versus equivalent strain rate}
	\end{subfigure}%
~
	\begin{subfigure}{0.5\textwidth}
	\includegraphics[trim={0.2cm 0.2cm 0.15cm 1cm}, clip, width=\textwidth]{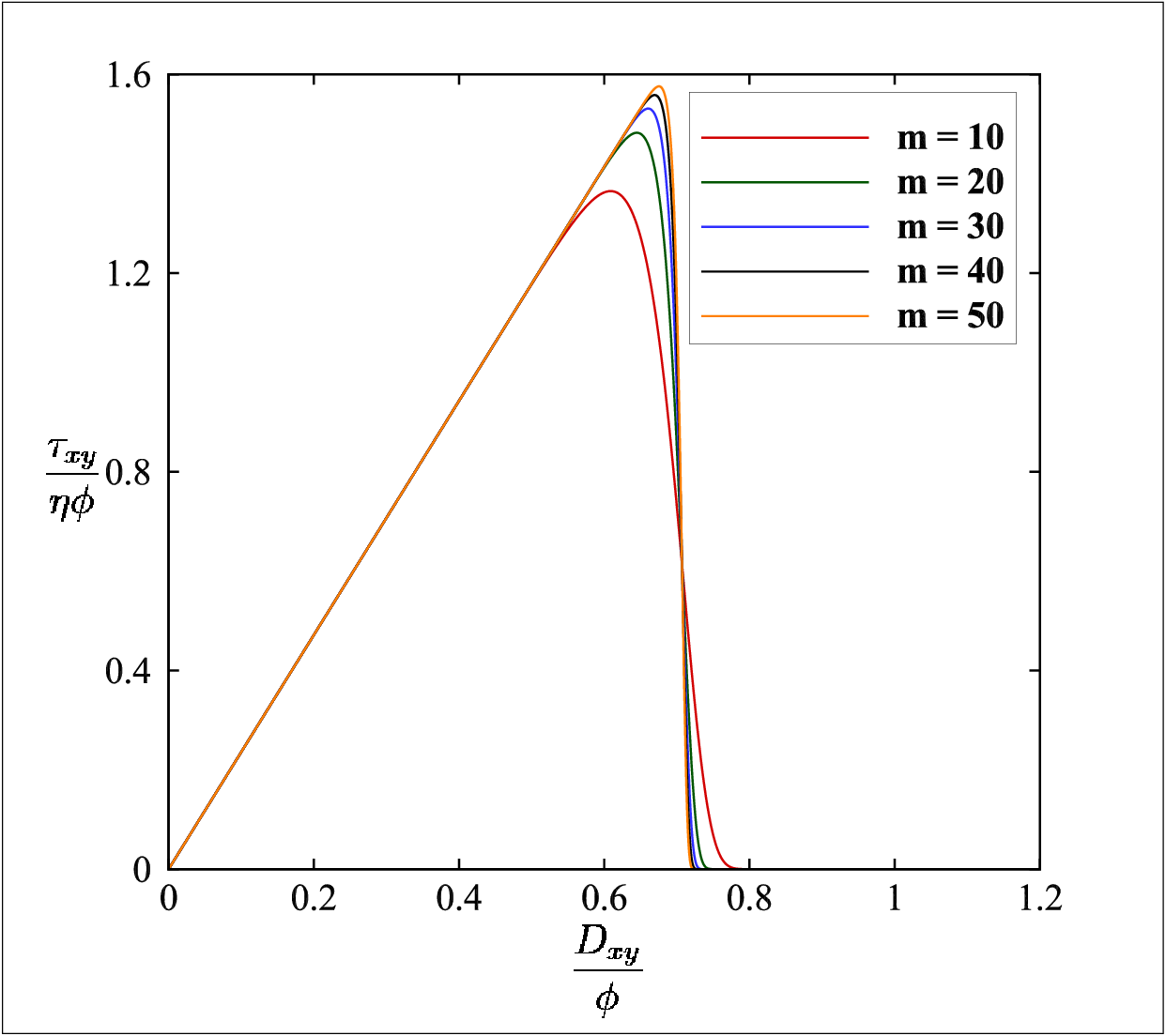}\vspace*{-0.1cm}
	\caption{\label{Fig_tau} stress versus strain rate}
	\end{subfigure}
\caption{Viscous strength model}
\end{figure}

By increasing the magnitude of constant $m$ in Eq. \ref{eq_eta_vsm}, it is possible to approach the step function that physically means a breakdown in the internal friction between fluid layers -- Fig. \ref{Fig_eta}.

Accordingly, the modified constitutive model incorporating the viscous strength becomes
\begin{equation}
\boldsymbol{\tau} = 2\eta^*\mathbf{D}.
\label{eq_tau_vsm}
\end{equation}

For the sake of clarification, let us consider shear flow with nonzero strain rate component $D_{xy}$ and stress $\tau_{xy}$. In this case, the constitutive law takes the form
\begin{equation}
\frac{\tau_{xy}}{\eta\phi}=2\left(\frac{D_{xy}}{\phi}\right) \text{exp} \left[-2^m\left(\frac{D_{xy}}{\phi}\right)^{2m}\right].
\label{eq_tau_vsm1}
\end{equation}

Its graphical representation in Fig. \ref{Fig_tau} shows that, unlike the classical NS theory, the shear stress possesses a finite limit -- strength -- indicating the breakdown of internal viscous friction. The strain rate corresponding to this stress bound can be obtained by solving $d\tau_{xy}/dD_{xy} = 0$ for Eq. \ref{eq_tau_vsm1}, which results in
\begin{equation}
\frac{D_{xy}}{\phi}=\left[\frac{1}{2^{m+1}~m}\right]^{\frac{1}{2m}}.
\label{eq_tau_vsm2}
\end{equation}
Hence, the viscous strength of the fluid becomes $\tau_{xy} = \sqrt{2}\eta \phi$ for $m \gg\ 1$.\\

\textbf{Remark 1}
We note that the conventional Navier-Stokes assumption of the perfectly intact viscous bonds is an extreme case of the viscous strength model with $\phi \to \infty$. In other words, the traditional NS model possesses infinitely large viscous strength. The latter means that the friction between adjacent fluid layers is always ideal and the viscous stress can increase infinitely with the increasing strain rate. Such property of the NS model is doubtful on physical grounds because no unbreakable materials really exist.

\textbf{Remark 2}
The model parameter $m$ helps in the mathematical smoothing of the step function. For materials with finite viscous stress, the critical strain rate can be obtained with the help of equality $\lim_{m\to\infty}\left[\frac{1}{2^{m+1}~m}\right]^{\frac{1}{2m}} = \frac{1}{\sqrt{2}}$.

\section{\label{sec:3}Pipe flow simulations}

\begin{figure}[!h]
\centering
	\begin{subfigure}{0.5\textwidth}
		\includegraphics[trim={0 -7cm 0 0}, clip, width=\textwidth]{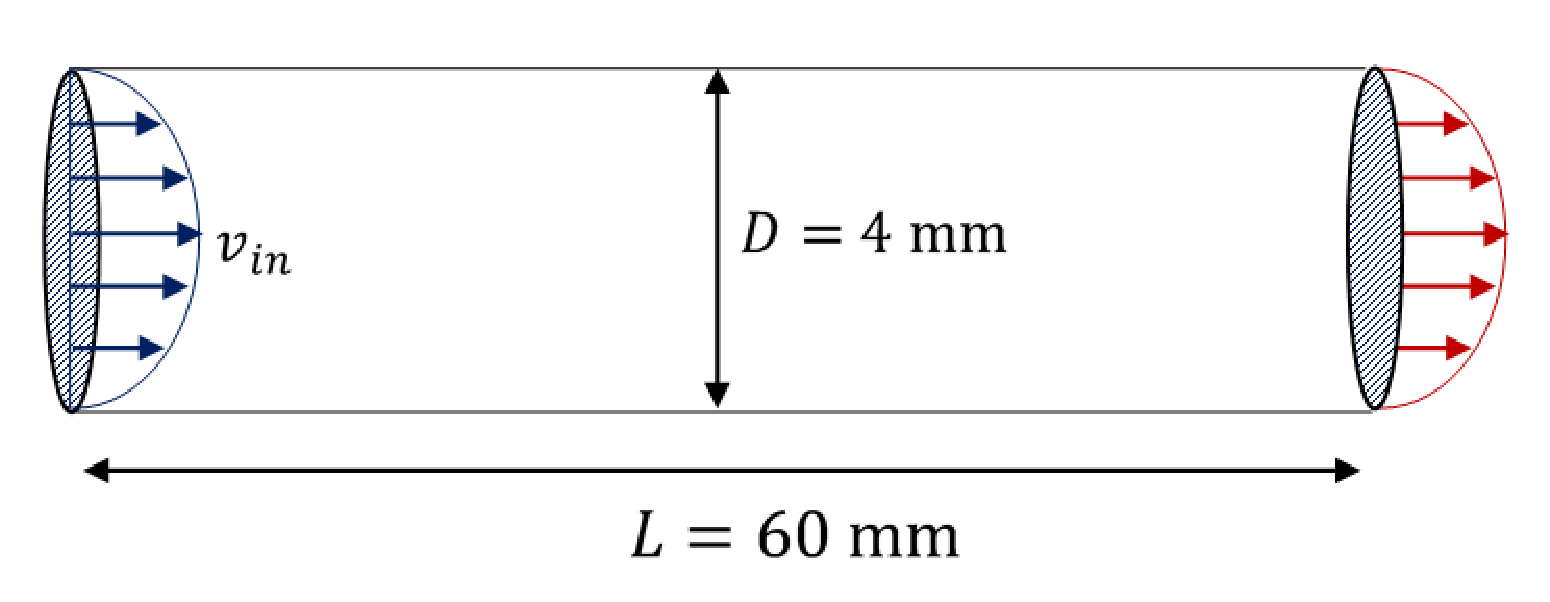}
		\caption{\label{Fig-geo} Three-dimensional flow through a cylindrical pipe}
	\end{subfigure}%
~
	\begin{subfigure}{0.5\textwidth}
		\includegraphics[trim={1cm 0 0 0}, clip, width=\textwidth]{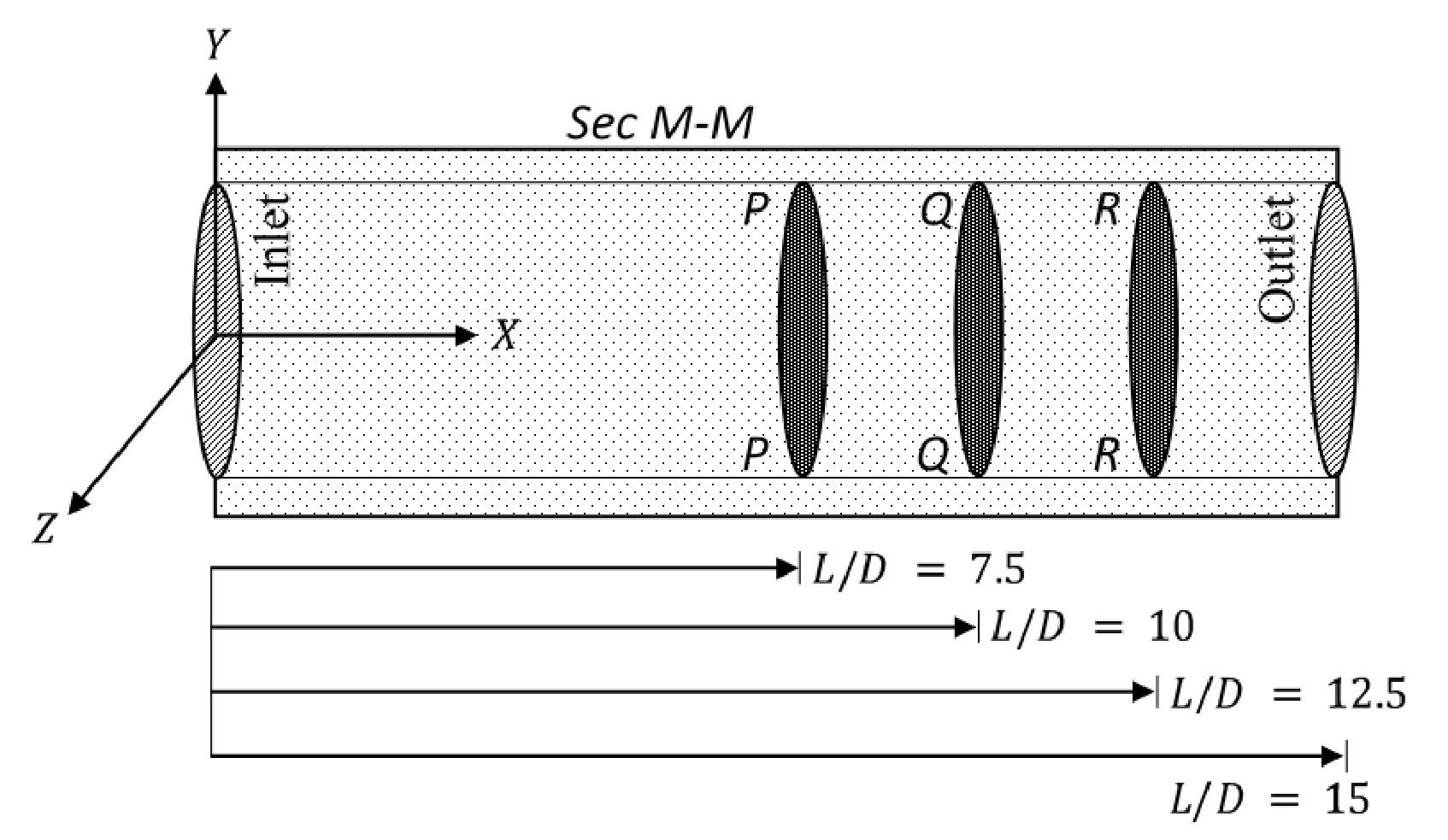}
		\caption{\label{Fig-sec} Section cut at specific locations in the pipe for recording and comparison of numerical results}
	\end{subfigure}
\caption{Schematic diagram of the problem}
\end{figure}

In this section, we describe simulations of a three-dimensional pipe flow through a 60 mm long perfectly straight cylindrical pipe with a uniform circular cross-section of diameter $D=4$ mm -- see Fig. \ref{Fig-geo}. We use two material models:

\begin{enumerate}
\item[(a)]{Conventional Navier-Stokes model with infinite viscous strength, further abbreviated CV};
\item[(b)]{Modified Navier-Stokes model with finite viscous strength, further abbreviated VSM}.
\end{enumerate}

The fluid is water with mass density $\rho = 1000~\text{kg}/\text{m}^3$, viscosity $\eta = 0.001~\text{kg}/\text{m}/\text{s}$, critical equivalent strain rate \cite{volokh2009investigation, volokh2013navier} $\phi = 848.5 ~\text{s}^{-1}$, and constant $m = 20$. For the sake of comparison, specific sections are chosen in the pipe (see Fig. \ref{Fig-sec}) to record measurements of respective simulations.

\begin{figure}[!h]
	\centering
	\includegraphics[trim={4cm 1cm 4cm 1cm}, clip, width=0.5\textwidth]{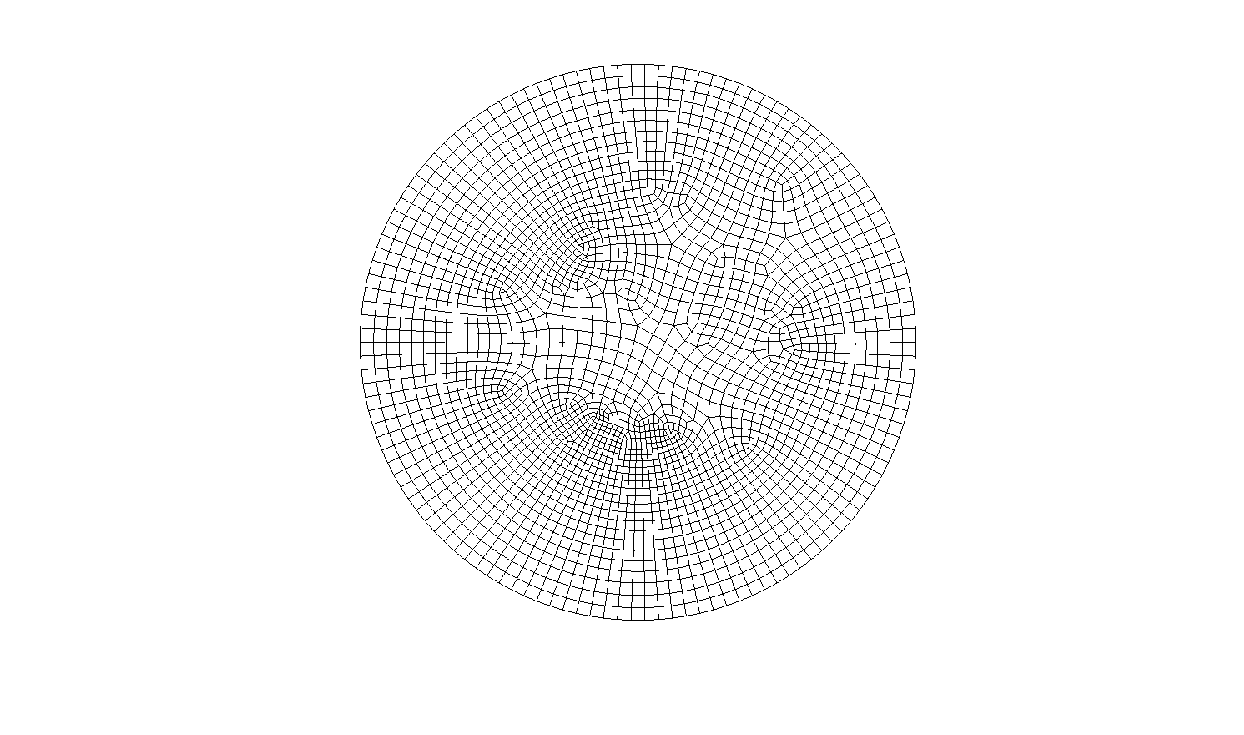}\vspace*{-0.1cm}
	\caption{\label{Fig-comp-mesh} Typical cross-sectional mesh}
\end{figure}

\begin{figure}[!h]
	\centering
	\includegraphics[trim={10cm 0 10cm 0}, clip, width=0.6\textwidth]{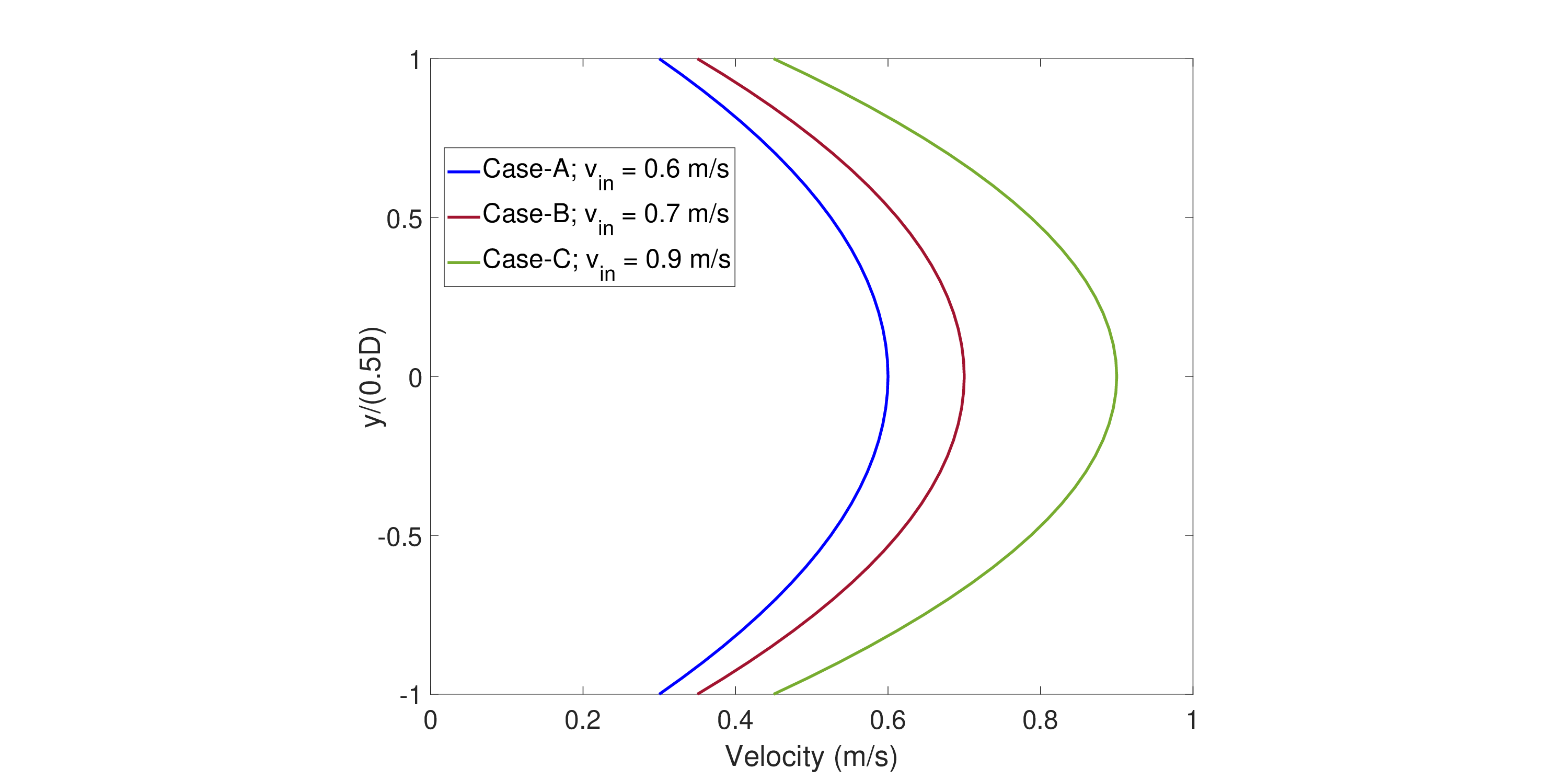}\vspace*{-0.1cm}
	\caption{\label{Fig-v_inlet} Velocity profiles provided at the inlet of the pipe for different simulation cases}
\end{figure}

We utilize commercial software ANSYS-2021R2 in computations. The spatial unstructured mesh has 1,476,706 elements with 1,514,133 nodes. The discretization at a typical cross-section of the pipe is shown in Fig. \ref{Fig-comp-mesh}. The wall of the pipe is stationary with no-slip boundary conditions. It is further assumed that the fluid flow is initially stable and, hence, a fully developed parabolic flow profile (Fig. \ref{Fig-v_inlet}) is provided at the inlet, and zero-pressure boundary conditions are incorporated at the outlet of the pipe.

Simulations are performed for different load cases by varying the peak inlet velocity $v_\text{in}$. By specifying the inlet velocity, it is easy to have a perception of the velocity gradient generated within the flow. An approximate measure of the Reynolds number can be assessed based on a constant velocity $\hat{v}$ that would generate the stable fully developed velocity profile at the inlet face as shown in Fig. \ref{Fig-v_inlet}. The relationship between $\hat{v}$ and $v_\text{in}$ can be obtained from the laminar flow theory as $\hat{v} = 2v_\text{in}/3$. Hence, the Reynolds number becomes $Re =2\rho v_\text{in} D/(3\eta)$. Table \ref{Load_Case} lists the inlet velocities and corresponding Reynolds numbers for all loading cases considered in numerical examples.
\begin{table}
\centering
\caption{\label{Load_Case} Inlet velocity ($v_\text{in}$) and Reynolds number ($Re$) for different load cases ($\rho=1000~\text{kg}/\text{m}^3$; $D=4~\text{mm}$; $\eta=0.001~\text{kg}/\text{m}/\text{s}$)}\vspace*{-0.1cm}
\begin{tabular}{ccccc}
\hline
Load case	& Inlet velocity ($v_{\text{in}}$) (m/s)	&  $Re = \dfrac{2\rho v_{\text{in}} D}{3 \eta}$	\\
\hline
A		& 0.6					& 1600											\\
B		& 0.7					& 1867											\\
C		& 0.9					& 2400											\\
\hline
\end{tabular}
\end{table}

The choice of an appropriate numerical framework is essential to capture the underlying physical processes governing the genesis of material instability in a pipe flow problem. For realistic 3D flows, the direct numerical solution becomes intractable. Instead, the large Eddy simulation (LES) is routinely used to save time. The specific advantage of LES lies in effectively reducing computational costs, yet capturing the transition to turbulence with reasonable accuracy. The LES filters out the small-scale eddies, insignificant to turbulence, from the solution and determines the unknowns by a sub-grid turbulence model. Among various sub-grid turbulence models \cite{smagorinsky1963general, germano1991dynamic, nicoud1999subgrid, lilly1992proposed, shur2008hybrid} reported in the literature, the wall adopted local eddy-viscosity (WALE) \cite{nicoud1999subgrid} approach is advantageous for problems related to transitional flows. In principle, the WALE model represents proper scaling at the wall functions and returns a zero turbulent eddy viscosity ($\mu_\text{t}$) for laminar shear flows, which makes it appropriate to treat laminar zones in the fluid domain.
\begin{figure}[!h]
	\centering
	\begin{subfigure}{0.65\textwidth}
		\includegraphics[trim={0 0 0 0}, clip, width=\textwidth]{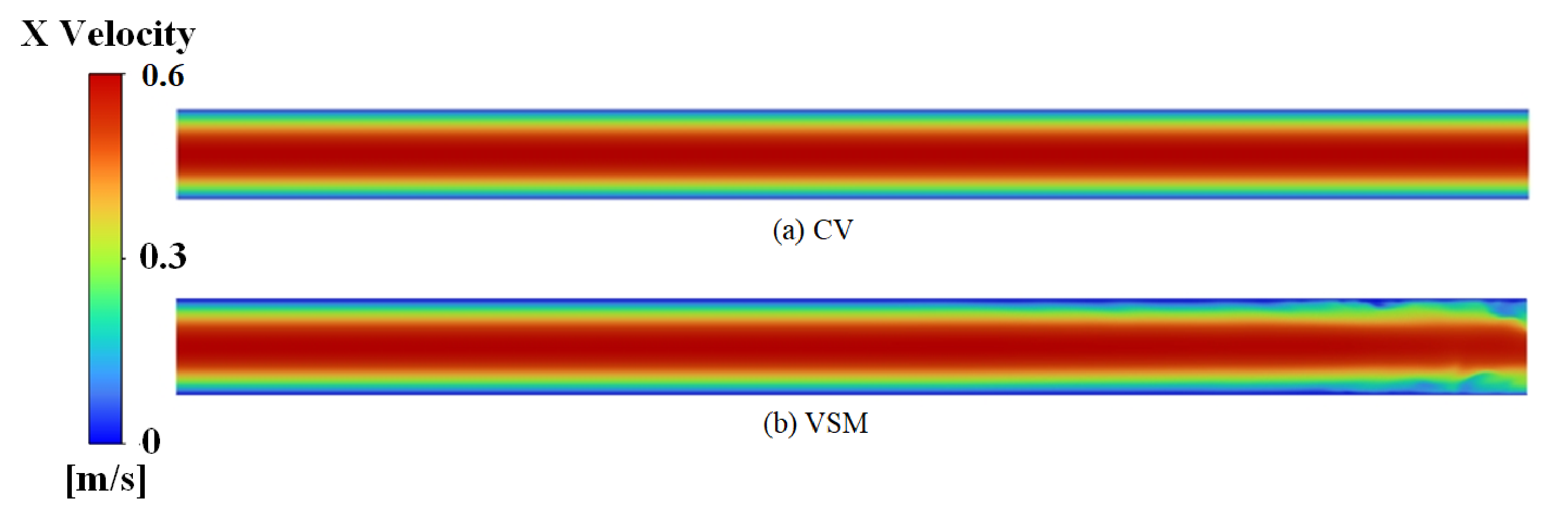}\vspace*{0.1cm}
		\caption{\label{Fig-vxcomp-06} Section - M-M (Case-A)}
	\end{subfigure}%
~
	\begin{subfigure}{0.35\textwidth}
		\includegraphics[trim={0 0 0 0}, clip, width=\textwidth]{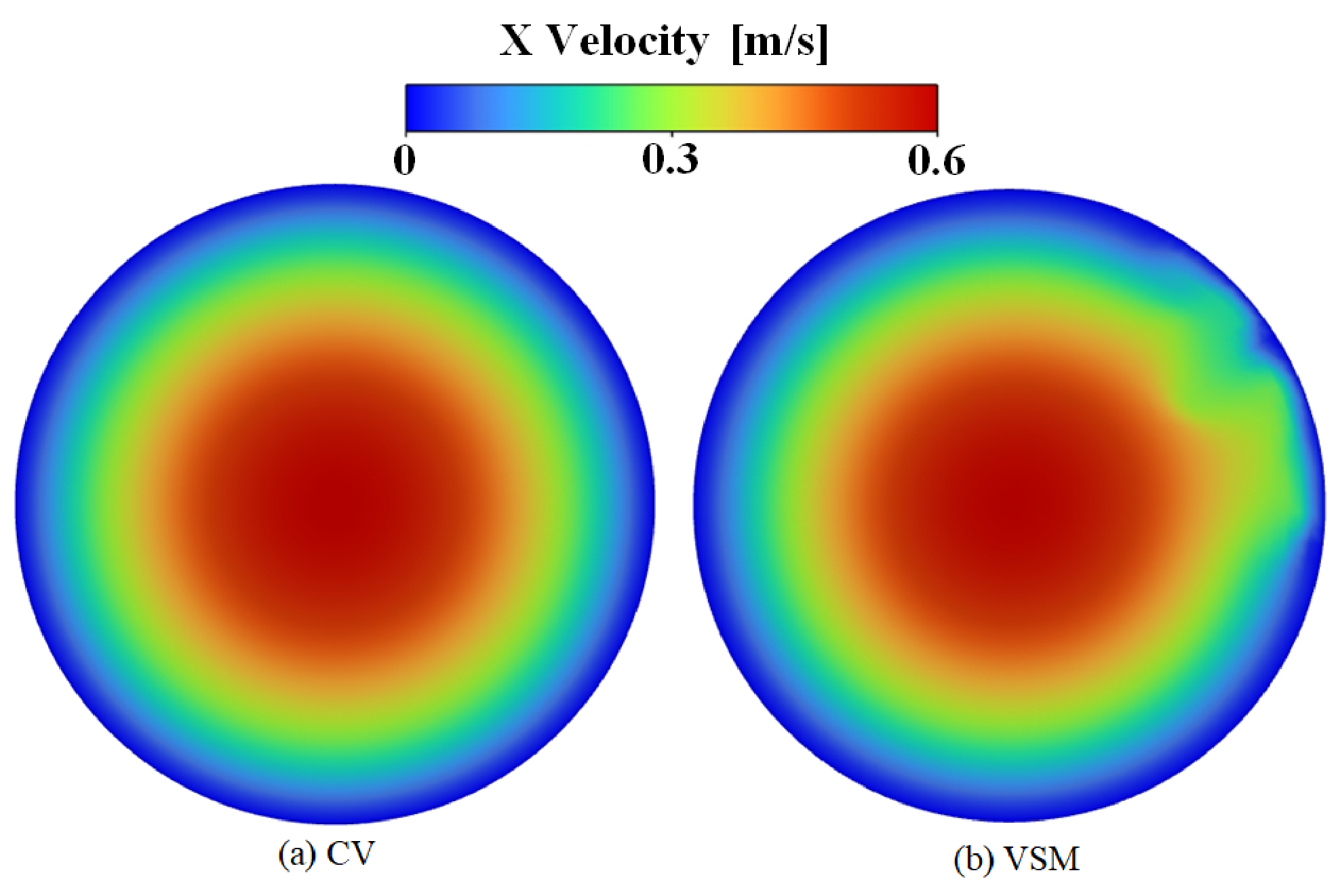}\vspace*{0.1cm}
		\caption{\label{Fig-vxcomp-10D06} Section - Q-Q (Case-A)} 
	\end{subfigure}%
\\
	\begin{subfigure}{0.65\textwidth}
		\includegraphics[trim={0 0 0 0}, clip, width=\textwidth]{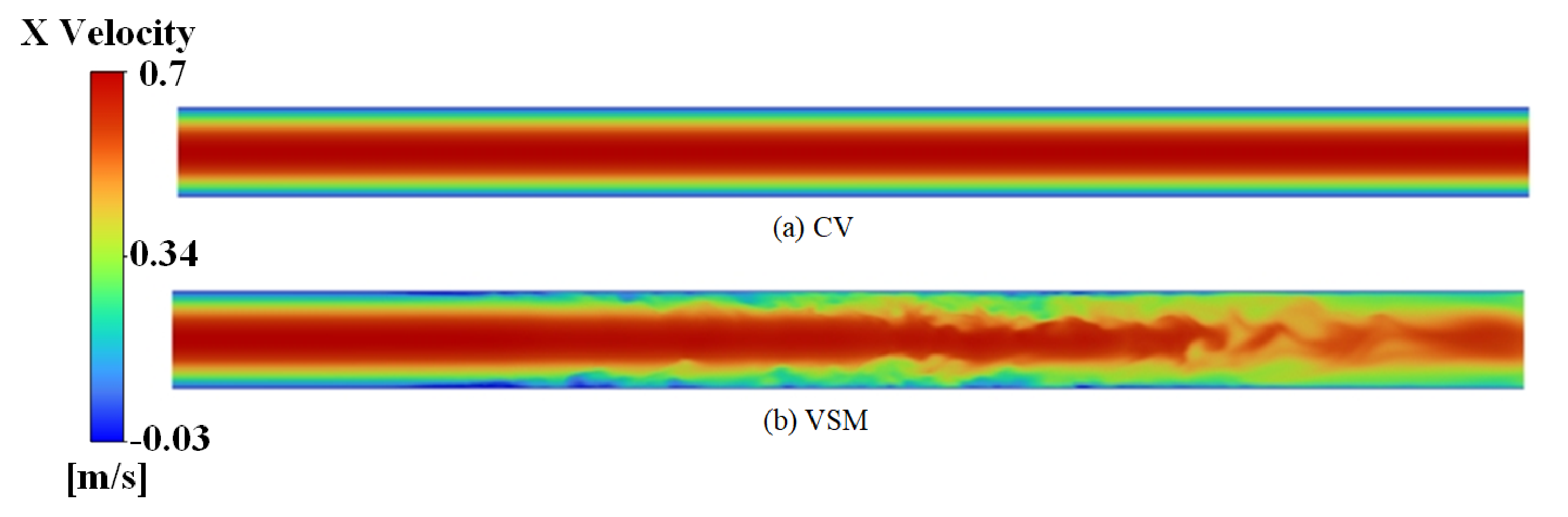}\vspace*{0.1cm}
		\caption{\label{Fig-vxcomp-07} Section - M-M (Case-B)}
	\end{subfigure}%
~
	\begin{subfigure}{0.35\textwidth}
		\includegraphics[trim={0 0 0 0}, clip, width=\textwidth]{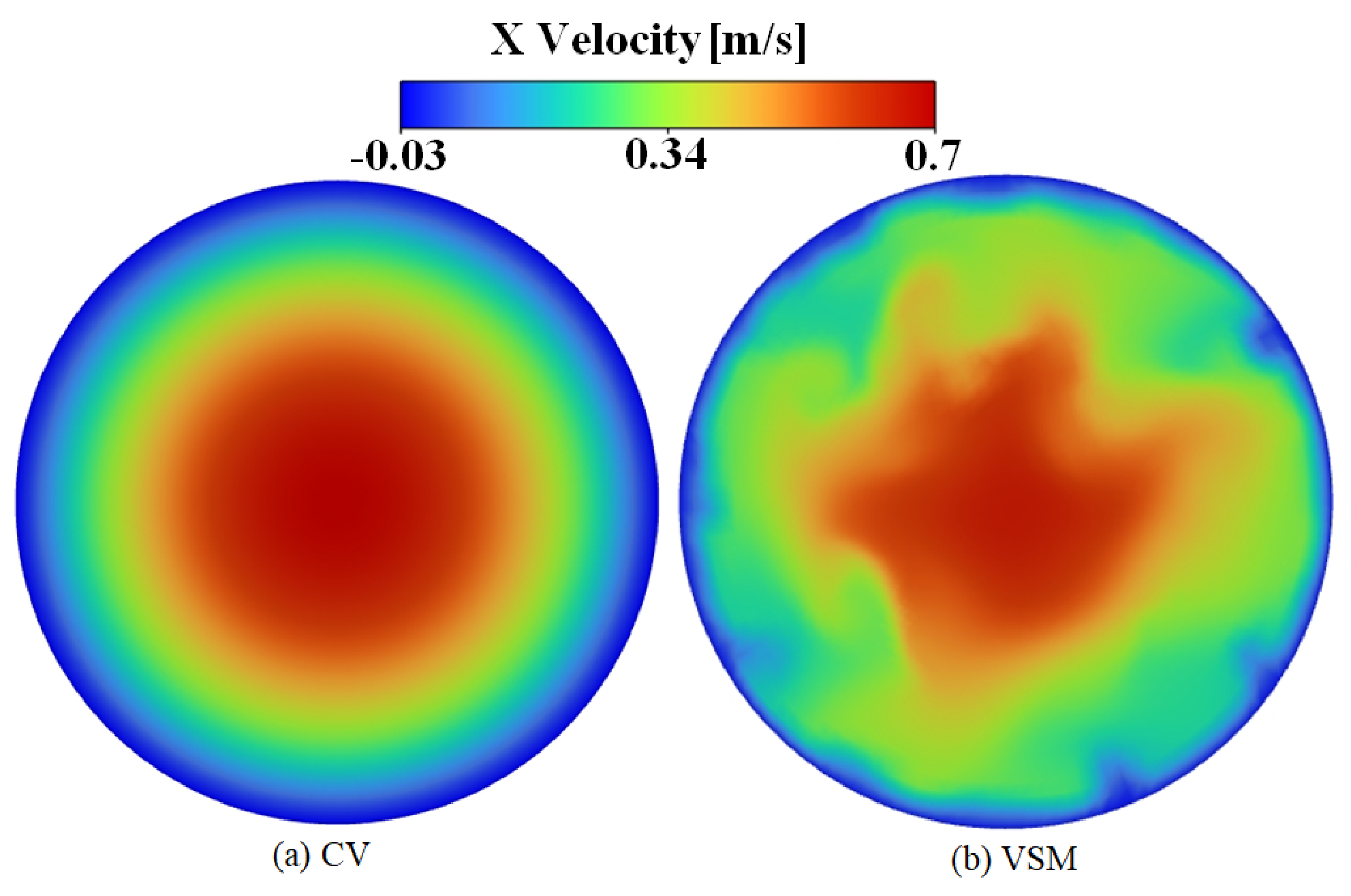}\vspace*{0.1cm}
		\caption{\label{Fig-vxcomp-10D07} Section - Q-Q (Case-B)} 
	\end{subfigure}%
\\
	\begin{subfigure}{0.65\textwidth}
		\includegraphics[trim={0 0 0 0}, clip, width=\textwidth]{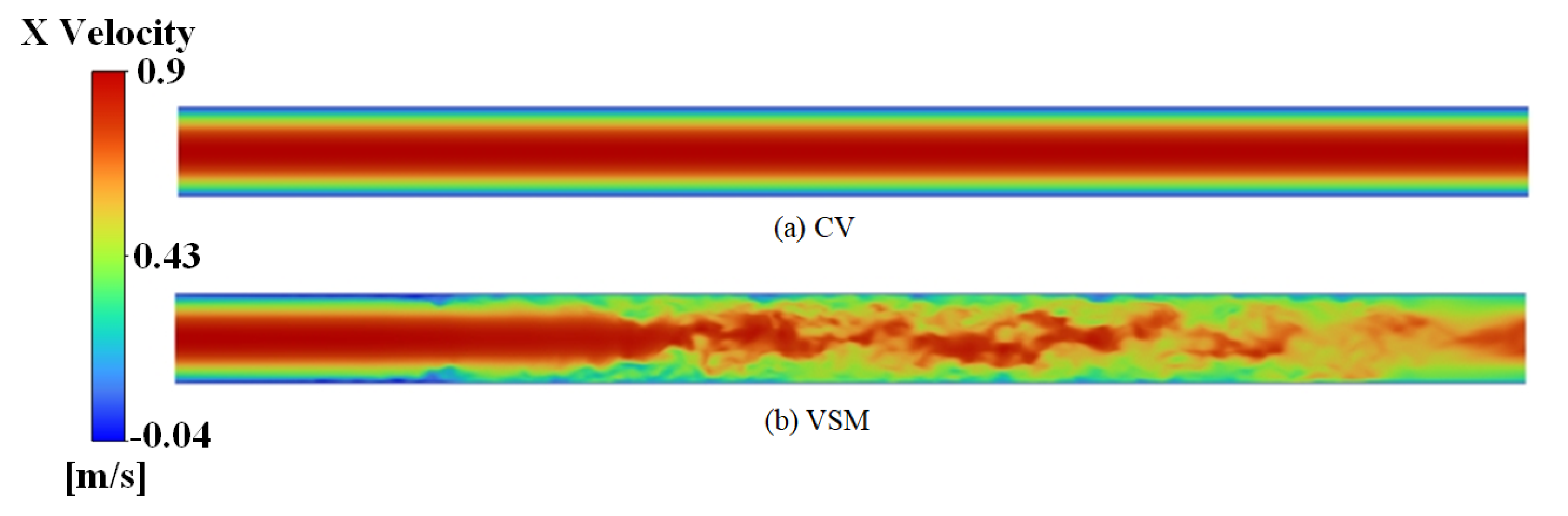}\vspace*{0.1cm}
		\caption{\label{Fig-vxcomp-09} Section - M-M (Case-C)}
	\end{subfigure}%
~
	\begin{subfigure}{0.35\textwidth}
		\includegraphics[trim={0 0 0 0}, clip, width=\textwidth]{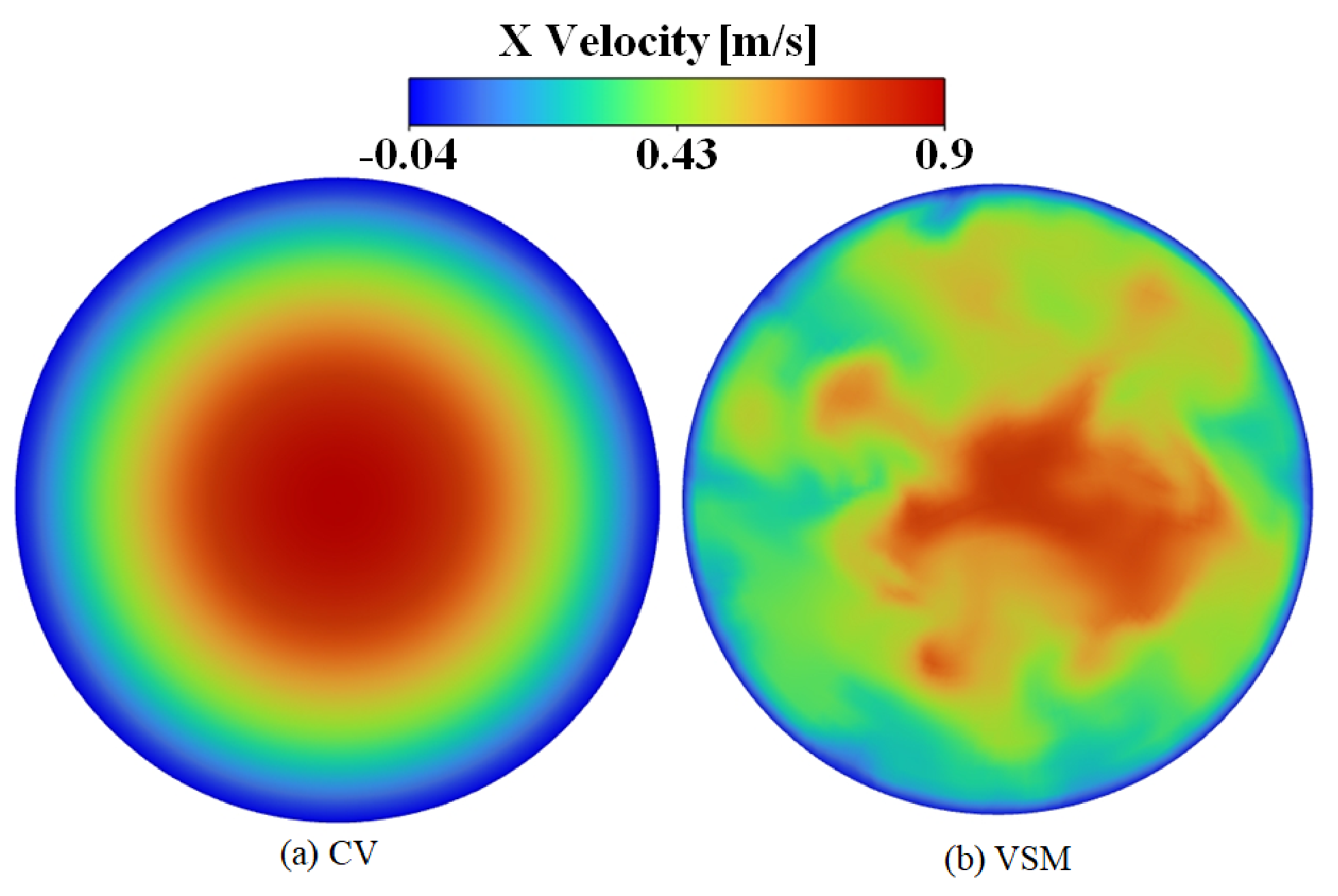}\vspace*{0.1cm}
		\caption{\label{Fig-vxcomp-10D09} Section - Q-Q (Case-C)} 
	\end{subfigure}%
\caption{Variation of flow velocity ($v_x$) (refer to sections M-M, and Q-Q in Fig.\ref{Fig-sec}) for different inlet velocity conditions --- A, B, \& C (Table \ref{Load_Case})}
\end{figure}

In the first set of simulations, with the peak inlet velocity of $v_{\text{in}} = 0.6$ m/s i.e. $Re \approx 1600$ (refer to Table \ref{Load_Case}) the CV model (i.e. the classical NS model with infinite viscous strength) predicts a stable laminar flow; whereas in the VSM (i.e. the modified NS model with finite viscous strength) we can observe instability in the flow. This highly localized material instability triggers deviation from laminar flow near the outlet of the pipe (Fig. \ref{Fig-vxcomp-06}). The onset of the instability is visible in the profile of the longitudinal velocity across the cross section of the pipe at a distance $L/D=10$ from the inlet face (section Q-Q in Fig. \ref{Fig-sec}) of the pipe (Fig. \ref{Fig-vxcomp-10D06}).

As we increase the inlet velocity to $v_\text{in}=0.7$ m/s i.e.,  $Re\approx 1867$, the CV model still predicts a perfectly stable solution and the flow remains perfectly laminar. However, in the VSM solution we can observe a visible deviation from the laminar flow pattern (Fig. \ref{Fig-vxcomp-07}). In this case, the instabilities are more pronounced than the first case. The contour of longitudinal velocity across the cross-section of the pipe at a distance $L=10D$ (section Q-Q in Fig. \ref{Fig-sec}) in Fig. \ref{Fig-vxcomp-10D07} demonstrates a clear transition to turbulence. 

With the further increase in the inlet velocity to $v_\text{in}=0.9$ m/s, that is, $Re\approx2400$, the VSM simulation produces a turbulent flow with eddies spread globally along the length of the pipe (Fig. \ref{Fig-vxcomp-09}) while the CV model still provides a stable laminar flow. Like in the previous examples, in section Q-Q, $L=10D$ distant from the inlet, Fig. \ref{Fig-vxcomp-10D09} indicates complete loss of the laminar mode in the longitudinal flow.

\begin{figure}[!h]
	\centering
	\begin{subfigure}{0.5\textwidth}
		\includegraphics[trim={0 0 0 0}, clip, width=\textwidth]{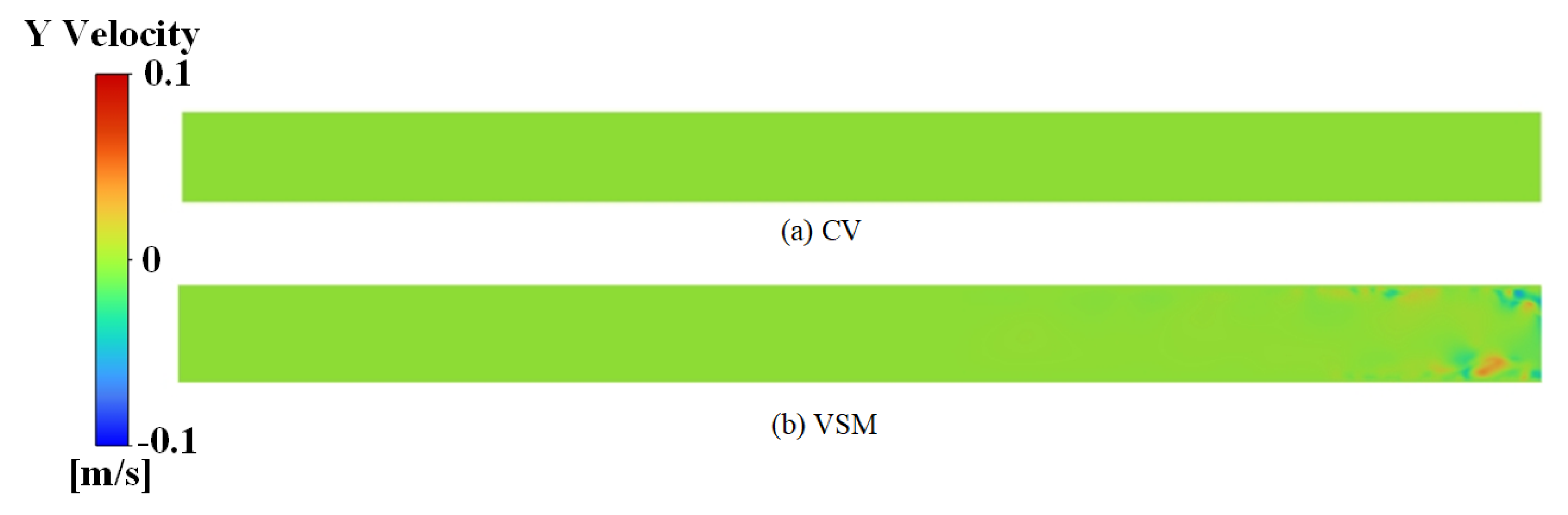}\vspace*{-0.1cm}
		\caption{\label{Fig-vycomp-06} Velocity in y-y direction $v_y$}
	\end{subfigure}%
~
	\begin{subfigure}{0.5\textwidth}
		\includegraphics[trim={0 0 0 0}, clip, width=\textwidth]{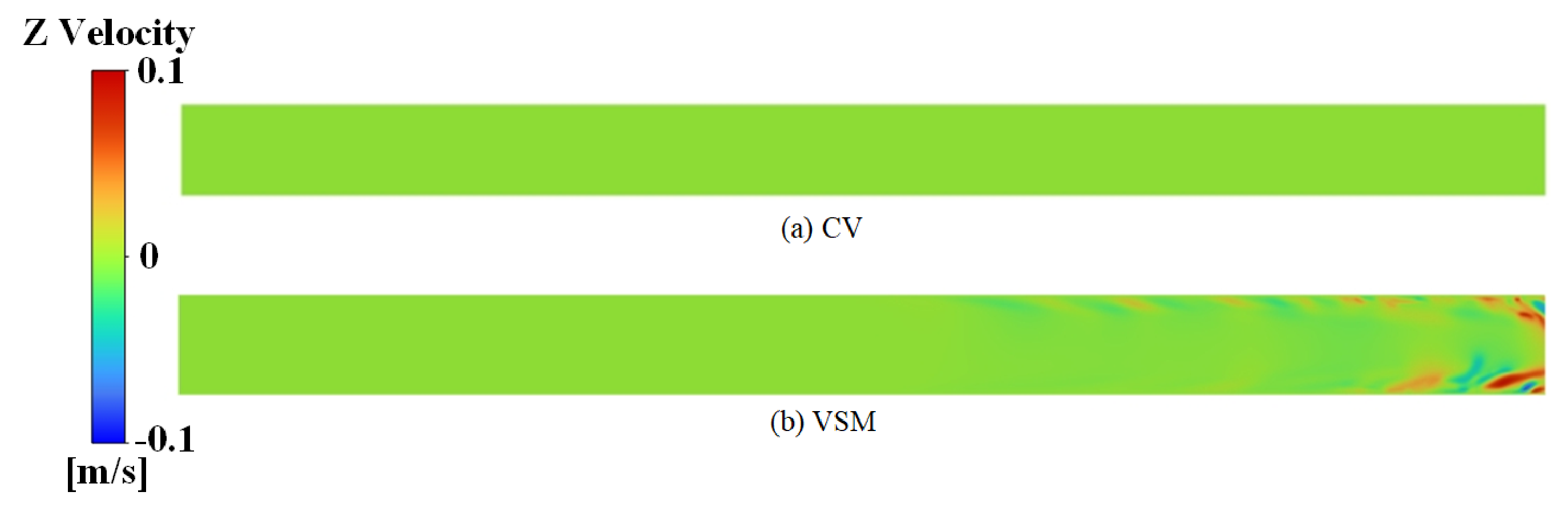}\vspace*{-0.1cm}
		\caption{\label{Fig-vzcomp-06} Velocity in the z-z direction $v_z$} 
	\end{subfigure}
\caption{Variation of transverse velocity along the length of the pipe at section M-M (refer to Fig.\ref{Fig-sec}) for inlet velocity case A ; $v_{\text{in}=0.6m/s}$}
\end{figure}
\begin{figure}[!h]
	\centering
	\begin{subfigure}{0.5\textwidth}
		\includegraphics[trim={0 0 0 0}, clip, width=\textwidth]{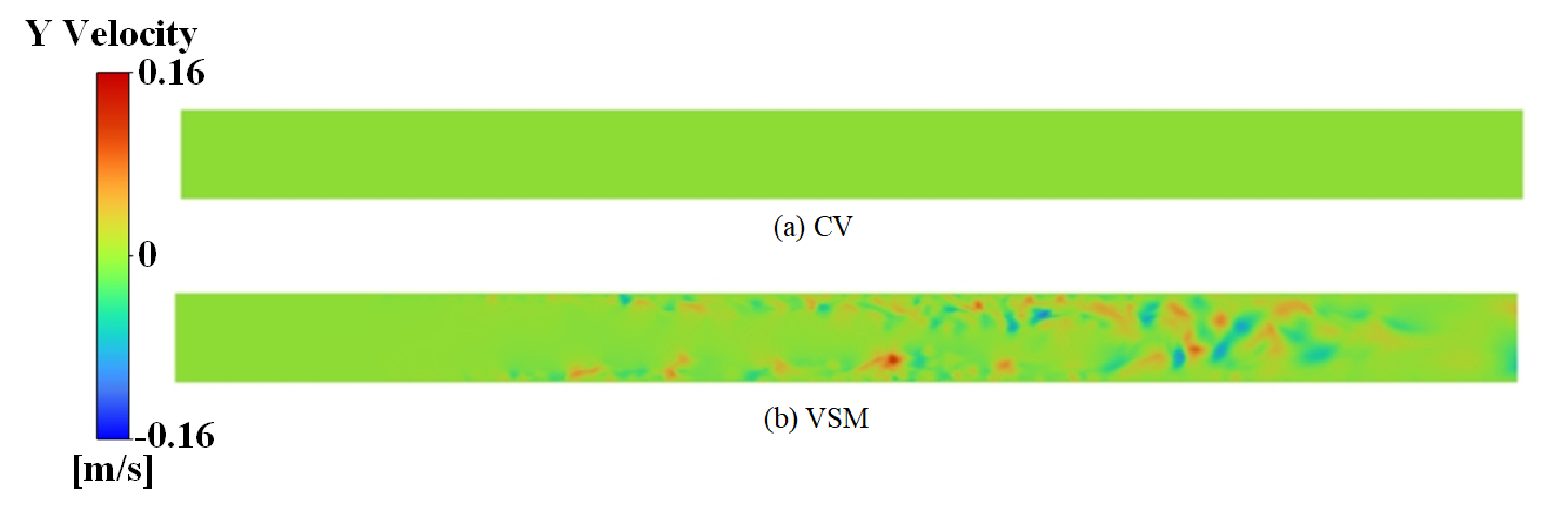}\vspace*{-0.1cm}
		\caption{\label{Fig-vycomp-07} Velocity in y-y direction $v_y$}
	\end{subfigure}%
~
	\begin{subfigure}{0.5\textwidth}
		\includegraphics[trim={0 0 0 0}, clip, width=\textwidth]{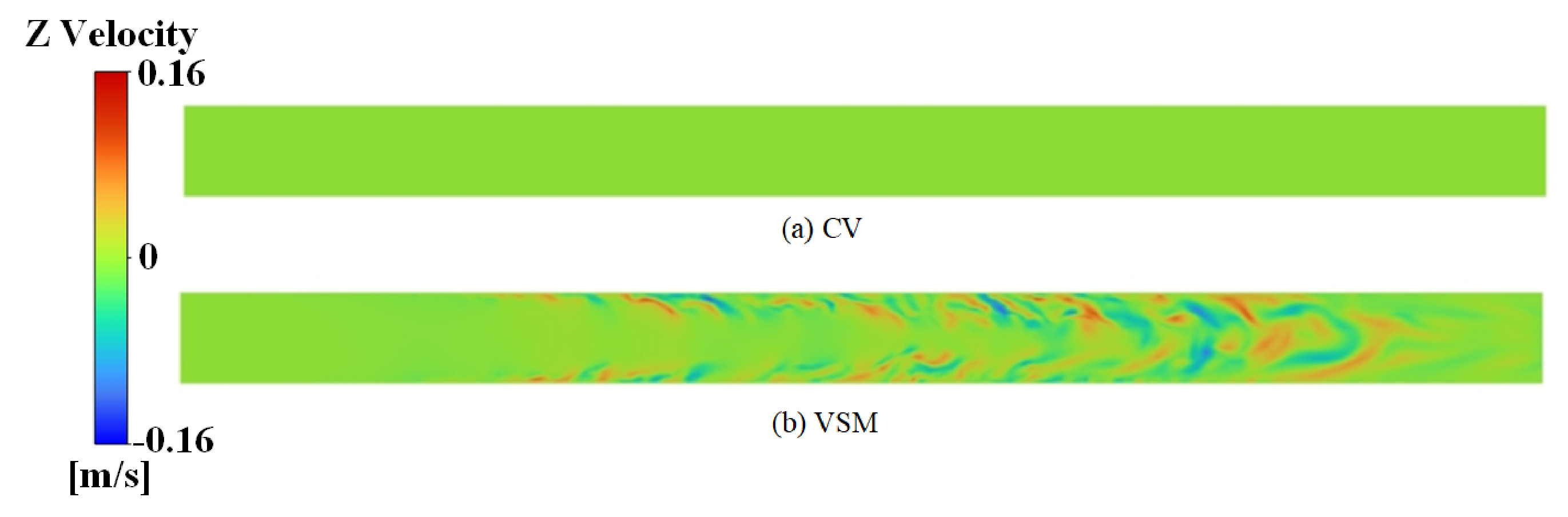}\vspace*{-0.1cm}
		\caption{\label{Fig-vzcomp-07} Velocity in the z-z direction $v_z$} 
	\end{subfigure}
\caption{Variation of transverse velocity along the length of the pipe at section M-M (refer to Fig.\ref{Fig-sec}) for inlet velocity case B ; $v_{\text{in}=0.7 m/s}$}
\end{figure}
\begin{figure}[!h]
	\centering
	\begin{subfigure}{0.5\textwidth}
		\includegraphics[trim={0 0 0 0}, clip, width=\textwidth]{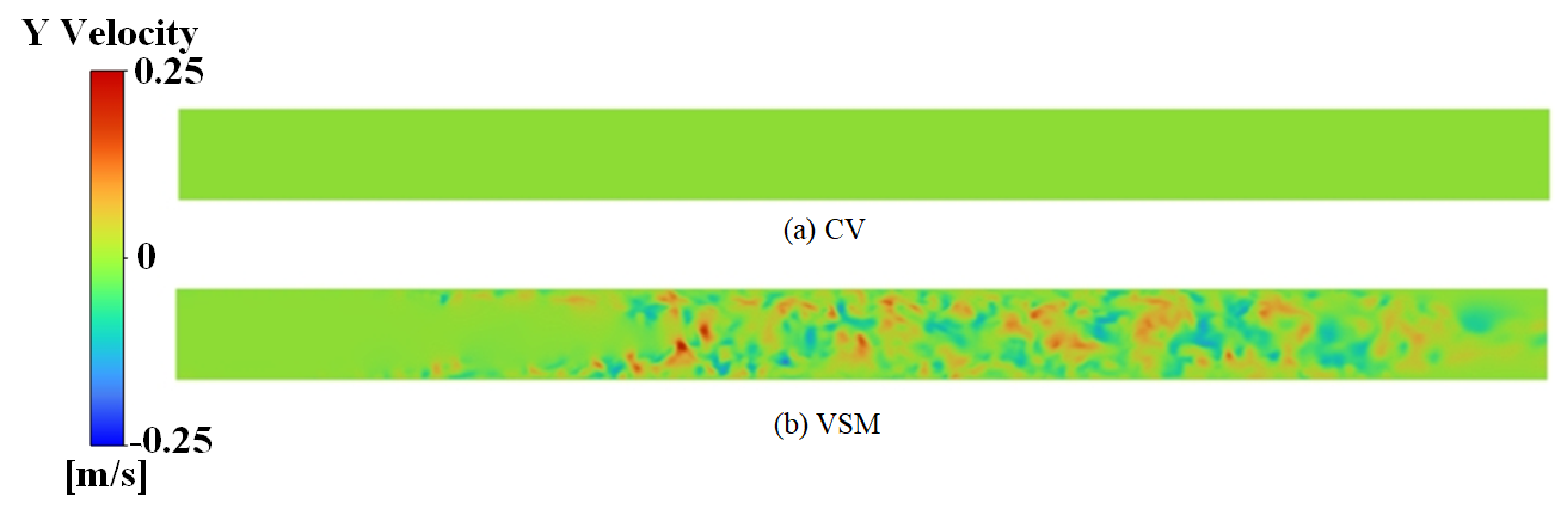}\vspace*{-0.1cm}
		\caption{\label{Fig-vycomp-09} Velocity in y-y direction $v_y$}
	\end{subfigure}%
~
	\begin{subfigure}{0.5\textwidth}
		\includegraphics[trim={0 0 0 0}, clip, width=\textwidth]{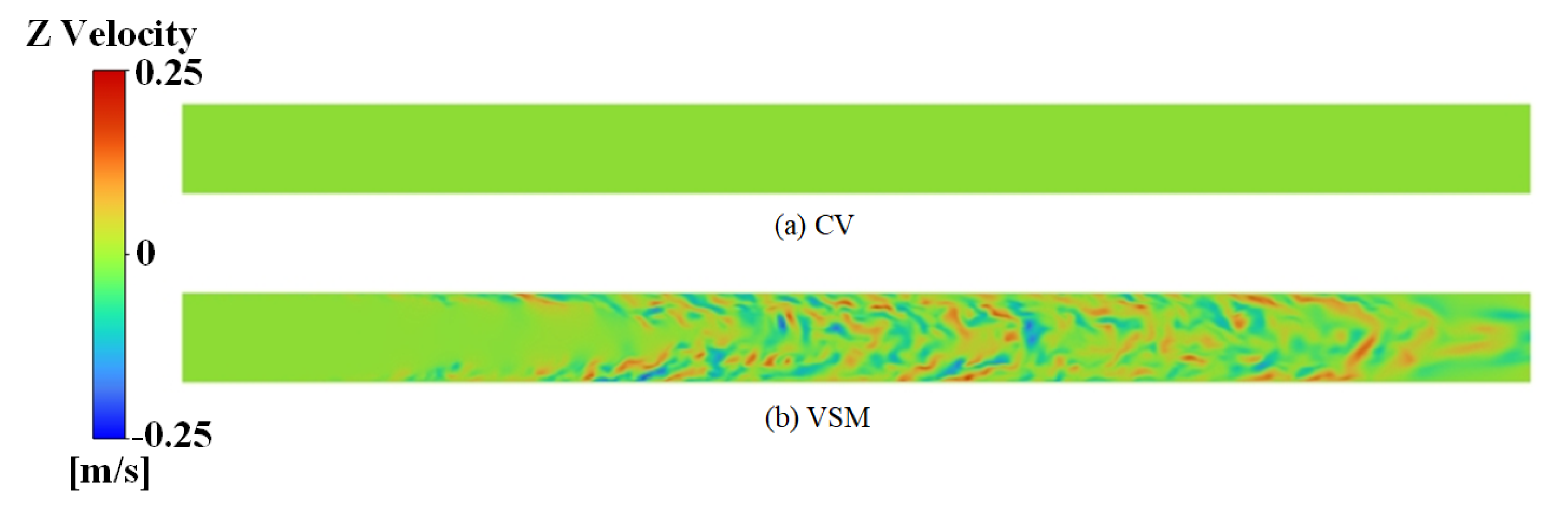}\vspace*{-0.1cm}
		\caption{\label{Fig-vzcomp-09} Velocity in the z-z direction $v_z$} 
	\end{subfigure}
\caption{Variation of transverse velocity along the length of the pipe at section M-M (refer to Fig.\ref{Fig-sec}) for inlet velocity case C ; $v_{\text{in}=0.9 m/s}$}
\end{figure}

Instabilities in the flow generates localised puffs or fluctuations in the transverse velocity profile in the channel. The VSM simulation reports that the peak fluctuation is $\approx 0.1$ m/s, whereas for the CV model we do not notice any significant fluctuation for $v_{\text{in} = 0.6}$ m/s (Figs. \ref{Fig-vycomp-06}-\ref{Fig-vzcomp-06}). These fluctuations are highly localized in nature and the flow remains laminar. As we increase the inlet velocity to $v_{\text{in} = 0.7}$ m/s the localised puffs magnify and the peak magnitude reaches $\approx 0.16$ m/s (Figs. \ref{Fig-vycomp-07}-\ref{Fig-vzcomp-07}). For $v_\text{in}=0.9$ m/s, that is, $Re \approx 2400$, the VSM simulation yields a turbulent flow with eddies with a maximum variation in the transverse velocity of $\approx 0.25$ m/s (refer to Figs. \ref{Fig-vycomp-09}-\ref{Fig-vzcomp-09}).\\

\textbf{Remark 3}
We note that the standard viscometer/rheometer tests are useless for the calibration of the fluid strength. Indeed, to observe the drop of viscosity, all fluid particles should reach the critical state simultaneously. That is practically impossible and the viscosity drops randomly at some points where the local material instability develops. Thus, only the observation of the onset of transition to chaos is suitable for calibration of the fluid strength. This issue is discussed with great attention in \cite{volokh2013navier}, to which the reader is referred. \\

\textbf{Remark 4}
We again emphasize that we used the same techniques of large eddy simulations for both the classical Navier-Stokes model and the model enhanced with strength. The latter model presents viscosity as a step function with non-zero and zero viscosity coefficients. The theory of large-eddy simulations is equally applicable to both constitutive models and the comparison of the results is objective.

\section{\label{sec:4}Discussion}
Based on the numerical simulations reported above, we suggest that the turbulence is the result of two main events: the genesis of material instabilities in the laminar flow and the transition and growth of local instabilities into large eddies. Instabilities appear initially due to the reduction in viscous stresses with the increase in flow strain rates.
\begin{table}[h]
\centering
\caption{\label{Tab_srt} Maximum strain rate and wall shear stress in the channel section for the CV model or conventional Navier-Stokes equation under different inlet velocity conditions ($\phi=848.5~\text{s}^{-1}$)}\vspace*{0.2cm}
\begin{tabular}{ccccc}
\hline
		& Inlet velocity & Max strain rate & Max wall shear &  \\
\hline
Case	& ($v_{\text{in}}$) & ($\sqrt{\mathbf{D}:\mathbf{D}}$)	& ($\tau_{\text{max}}$) & $\dfrac{\sqrt{\mathbf{D}:\mathbf{D}}}{\phi}$ \\
		& m/s &	$\text{s}^{-1}$	& Pa &  \\
\hline\\
A		& 0.6	& 596.7		& 0.60	& 0.70	\\
B		& 0.7	& 696.2		& 0.71	& 0.82	\\
C		& 0.9	& 895.2		& 0.91	& 1.06	\\
\hline
\end{tabular}
\end{table}

\begin{figure}[!h]
\centering
	\begin{subfigure}{0.5\textwidth}
		\includegraphics[trim={10.5cm 0 10.5cm 0}, clip, width=\textwidth]{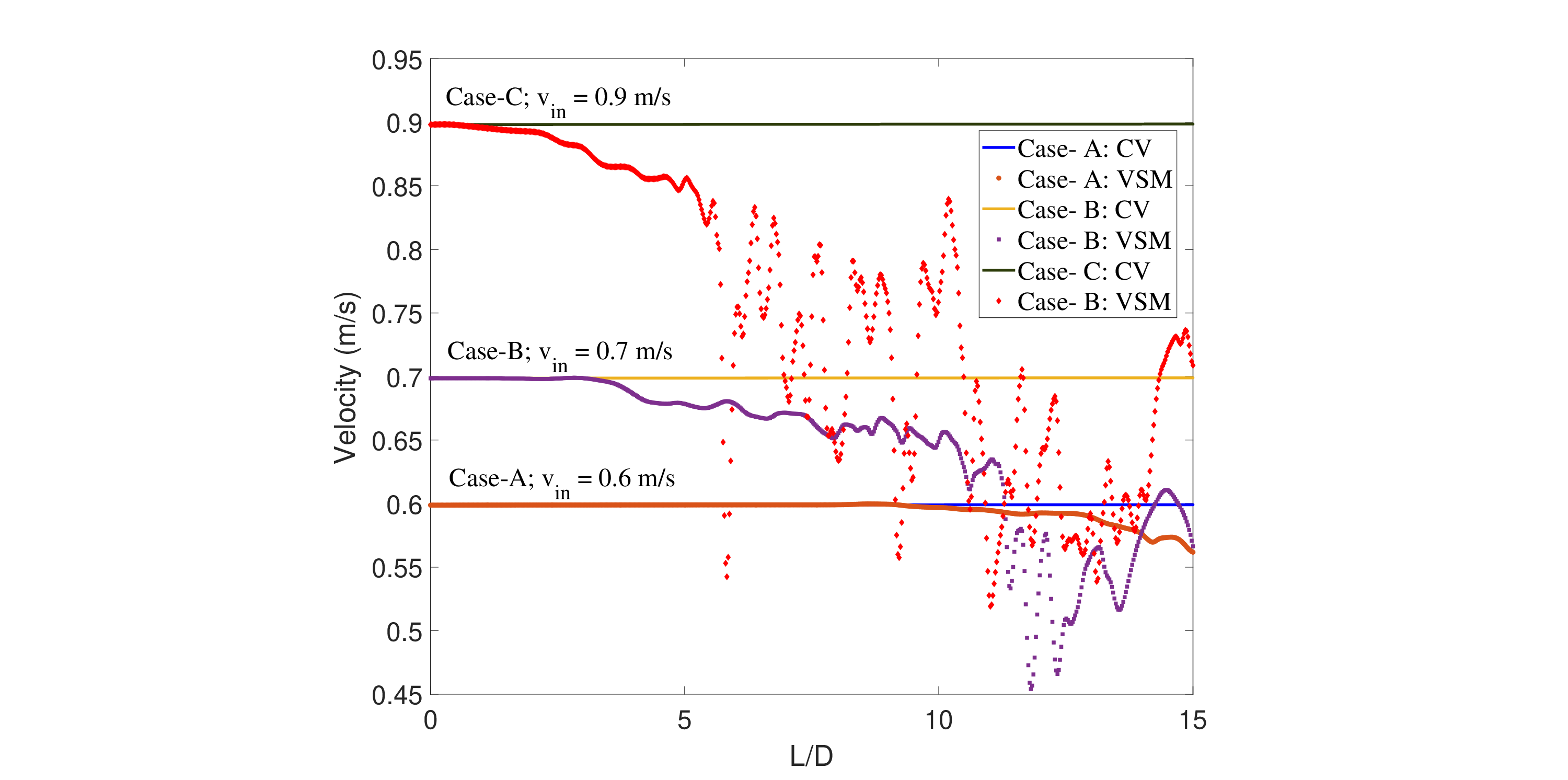}\vspace*{-0.1cm}
		\caption{\label{Fig-vx-XY} Profile of longitudinal velocity $v_x$}
	\end{subfigure}%
\\
	\begin{subfigure}{0.5\textwidth}
		\includegraphics[trim={10.5cm 0 10.5cm 0}, clip, width=\textwidth]{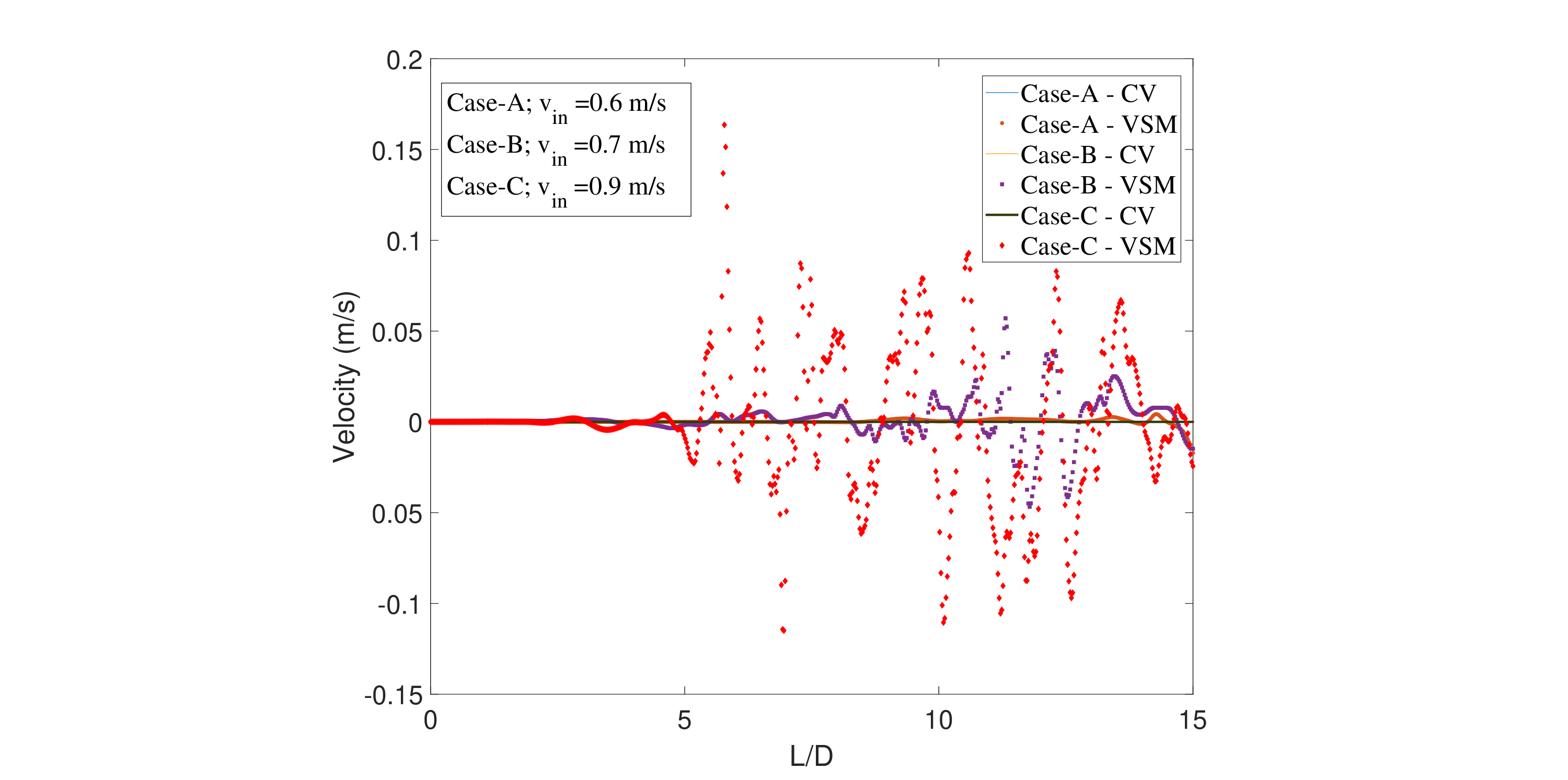}\vspace*{-0.1cm}
		\caption{\label{Fig-vy-XY} Profile of transverse velocity $v_y$} 
	\end{subfigure}%
~
	\begin{subfigure}{0.5\textwidth}
		\includegraphics[trim={10.5cm 0 10.5cm 0}, clip, width=\textwidth]{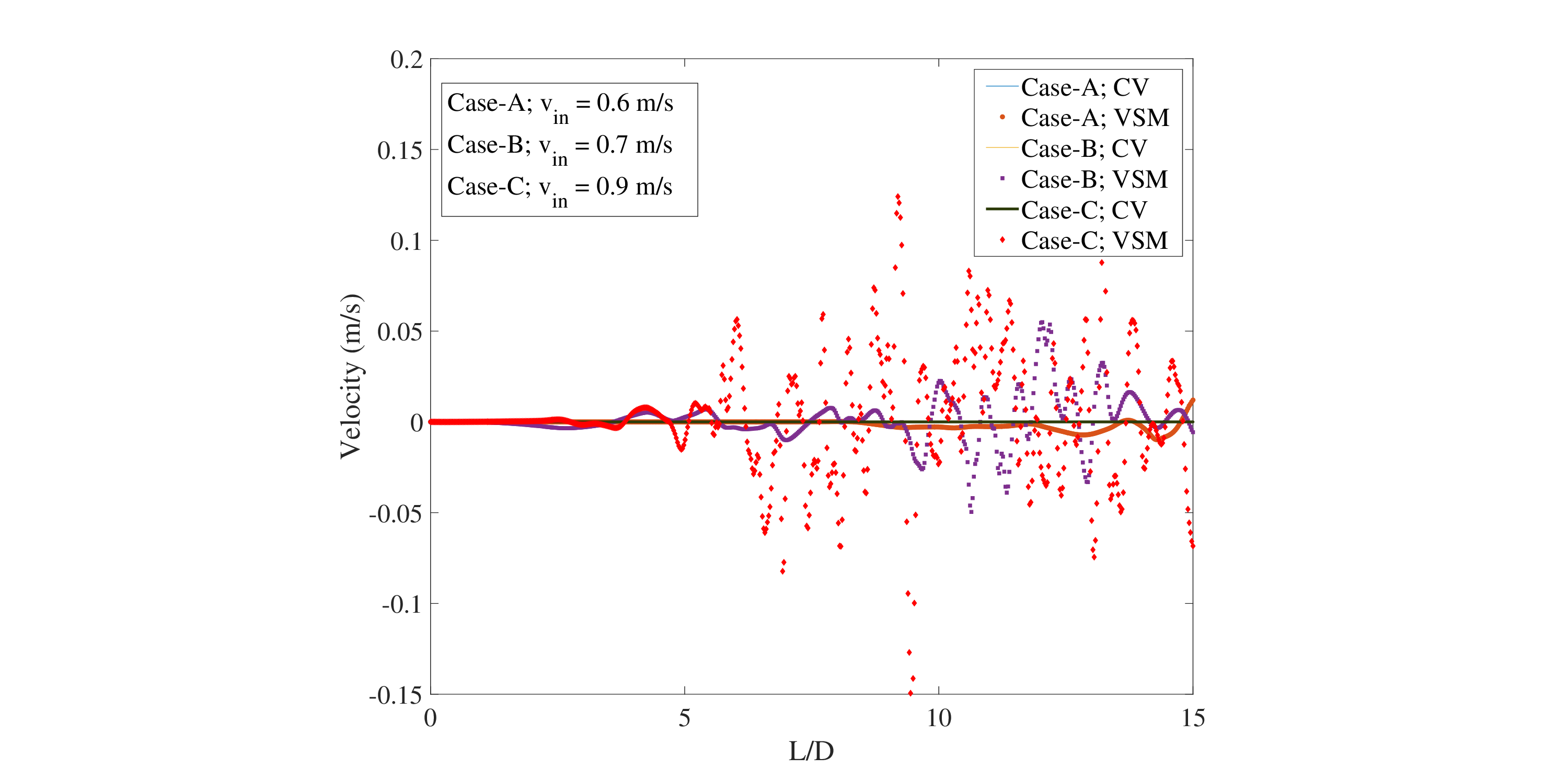}\vspace*{-0.1cm}
		\caption{\label{Fig-vz-XY} Profile of transverse velocity $v_z$}
	\end{subfigure}
\caption{Variation in flow velocities along the direction of pipe flow at section M-M (cut at the midplane where $z=0$, refer to Fig. \ref{Fig-sec}) for velocity cases A, B, and C} 
\end{figure}

Our simulations and some parameters collected in Table \ref{Tab_srt} prompt that material instabilities are born at $v_\text{in} = 0.6~\text{m/s}$. The latter means that shear stresses begin to drop near the pipe wall, small eddies appear and start to grow, and the flow deviates from the laminar (Fig. \ref{Fig-vx-XY}; Case-A) behavior. When the flow velocity increases to $v_\text{in}=0.7~\text{m/s}$, the fluid remains within the transition range, but the instabilities and eddies expand over a wider zone (Fig. \ref{Fig-vx-XY}; Case-B). By increasing the inlet velocity to $v_\text{in}=0.9~\text{m/s}$ (Case-C), the eddies become large and spread throughout the pipe (Fig. \ref{Fig-vx-XY}; Case-C).

Once the instability arises, the drop in the shear stress near the zone of instability induces disturbances in transverse velocities. The magnitude of the disturbances increases with increasing flow velocity (Figs. \ref{Fig-vy-XY}-\ref{Fig-vz-XY}). The turbulence moves towards the inlet face of the pipe with an increase in the inlet velocity of the fluid. The variation in axial velocity in section M-M is marginal for $v_\text{in} = 0.6~\text{m/s}$.
\begin{figure}[h]
	\centering
		\includegraphics[trim={0 0 0 0}, width=0.75\textwidth]{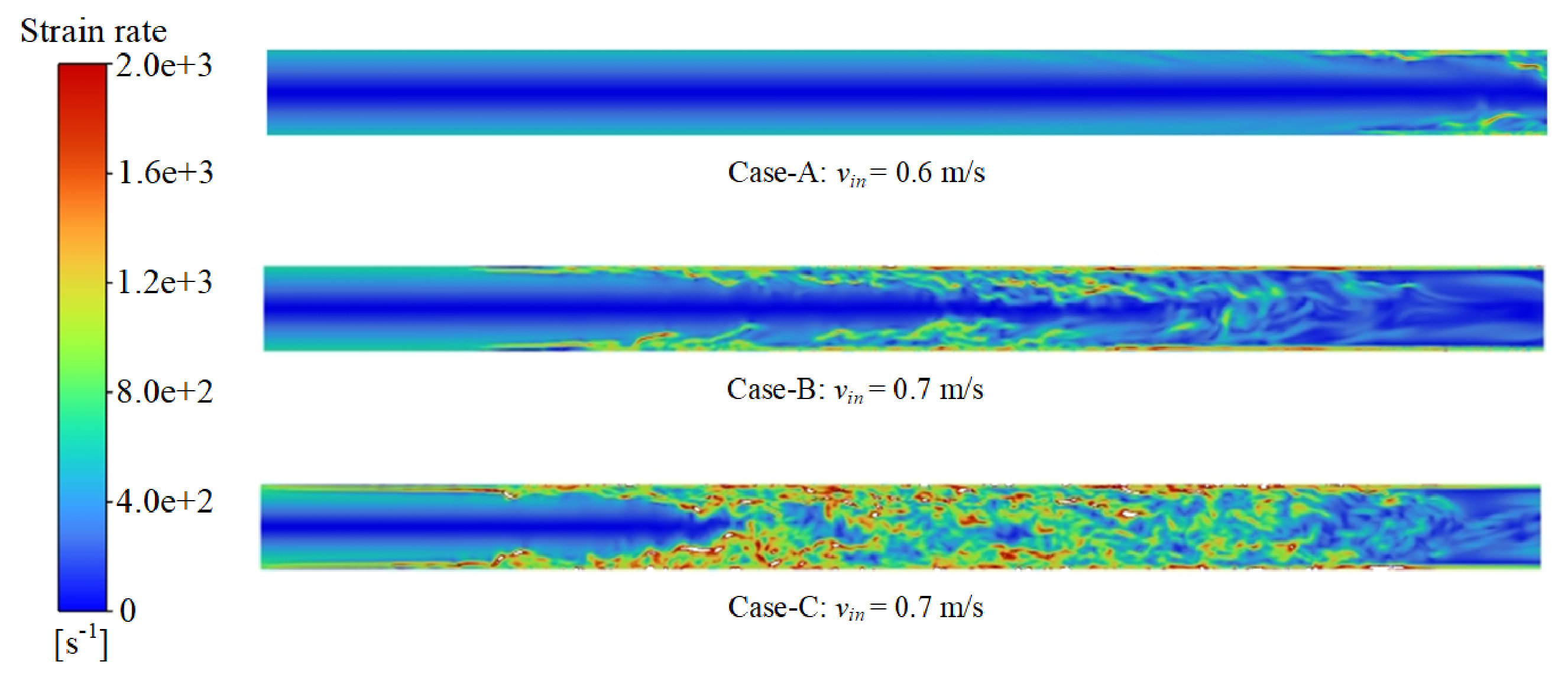}\vspace*{-0.1cm}
		\caption{\label{Fig-srtcomp} Magnification of the strain rate ($\sqrt{\mathbf{D}:\mathbf{D}}$) in the M-M section (at $z=0$, in Fig. \ref{Fig-sec}) for different inlet velocity conditions --- Case -- A,B, \& C}
\end{figure}

Fig. \ref{Fig-srtcomp} presents another interesting phenomenon: as the fluid loses its stability, the strain rates are magnified near the zone of instability. Subsequently, during the transition to turbulence, the magnified strain rates start propagating towards the center of the pipe. 

\begin{figure}[!h]
	\centering
	\begin{subfigure}{0.45\textwidth}
		\includegraphics[trim={10.5cm 0 12cm 0}, clip, width=\textwidth]{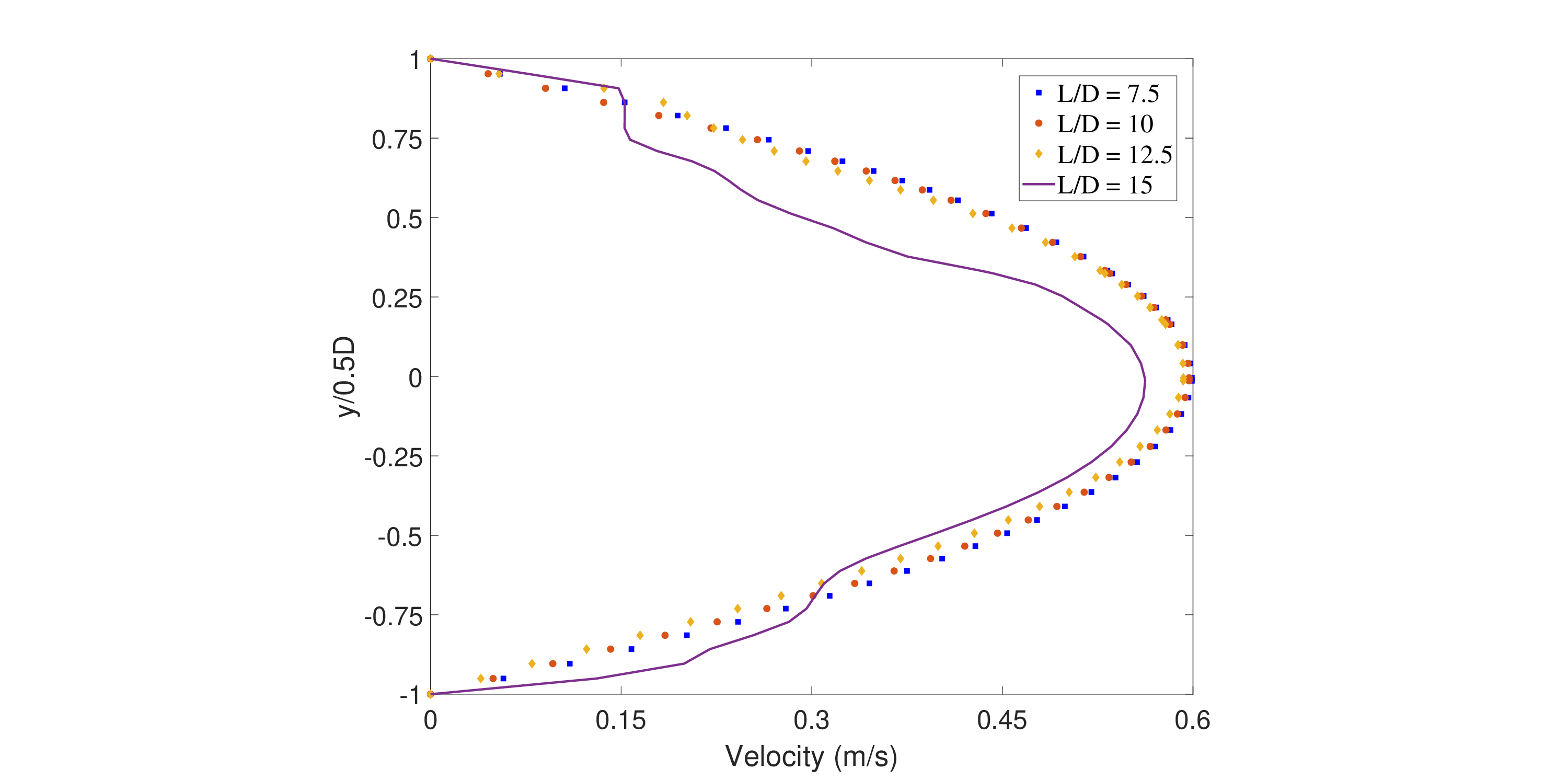}\vspace*{-0.1cm}
		\caption{\label{Fig-vx-06} $v_\text{in}=0.6$ m/s (Case-A)}
	\end{subfigure}%
~
	\begin{subfigure}{0.45\textwidth}
		\includegraphics[trim={10.5cm 0 12cm 0}, clip, width=\textwidth]{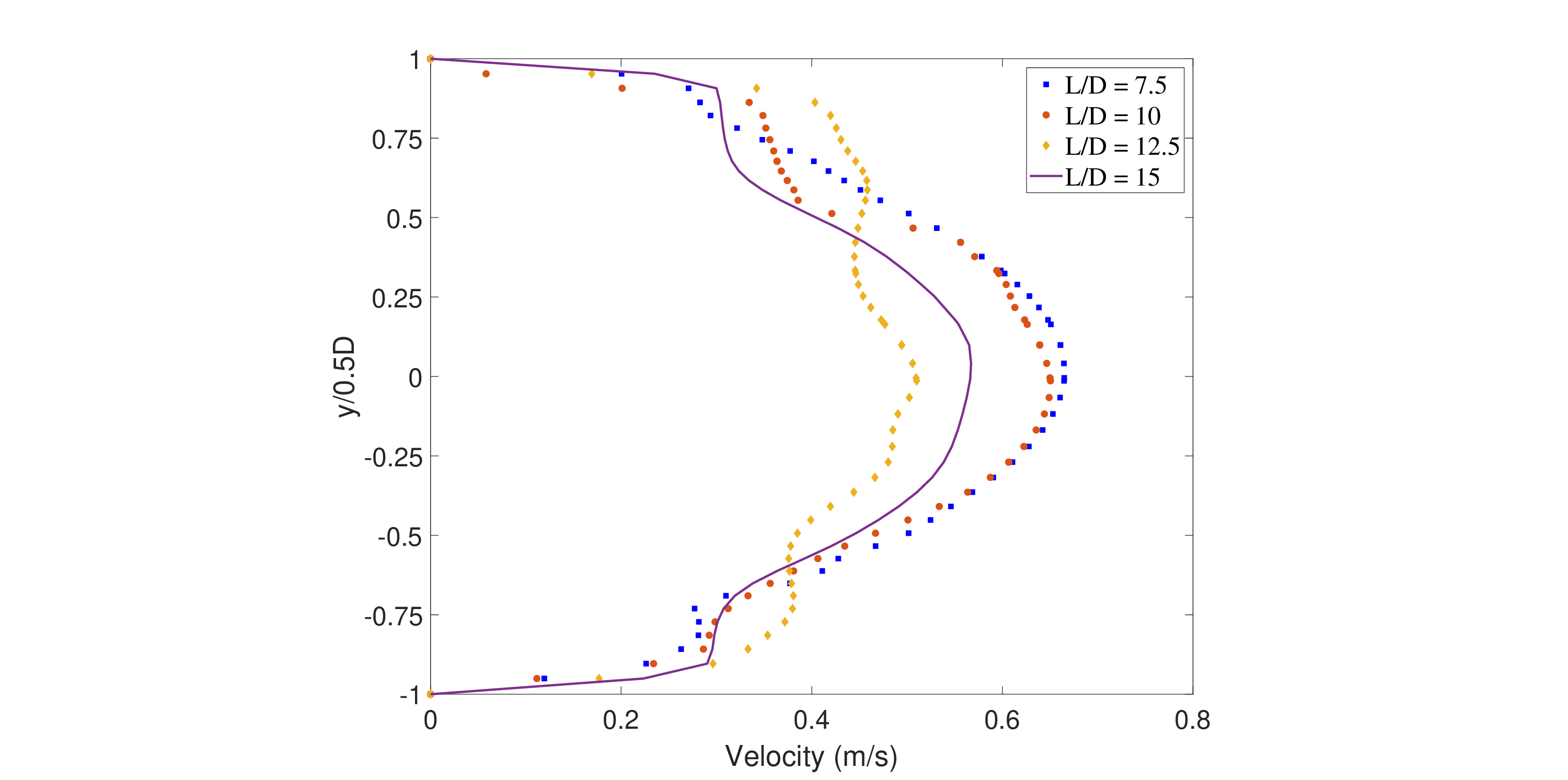}\vspace*{-0.1cm}
		\caption{\label{Fig-vx-07} $v_\text{in}=0.7$ m/s (Case-B)} 
	\end{subfigure}%
\\
	\begin{subfigure}{0.45\textwidth}
		\includegraphics[trim={10.5cm 0 12cm 0}, clip, width=\textwidth]{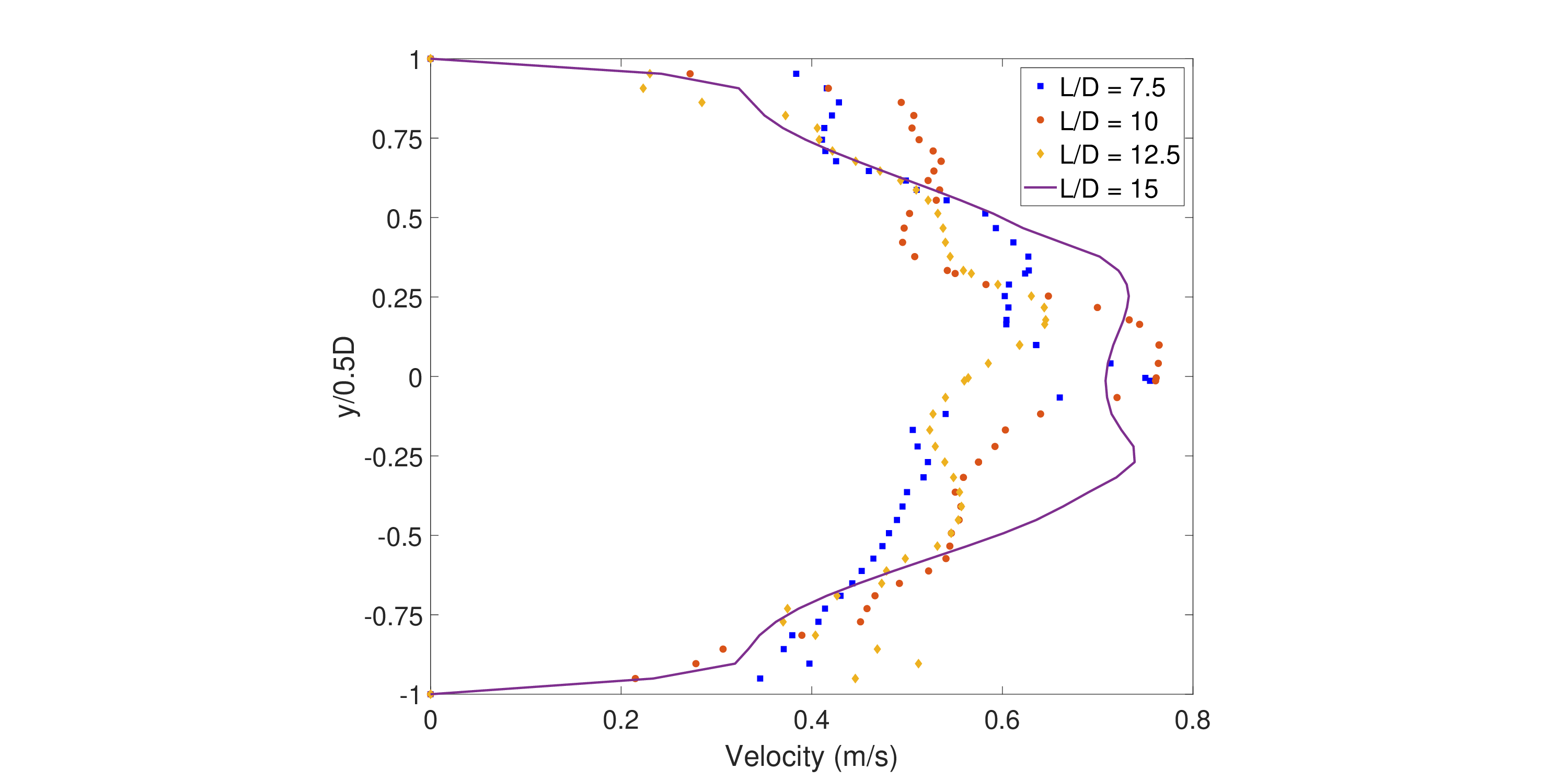}\vspace*{-0.1cm}
		\caption{\label{Fig-vx-09} $v_\text{in}=0.9$ m/s (Case-C)}
	\end{subfigure}
\caption{Variation of longitudinal velocity ($v_x$) across the cross-section of the pipe cut at distances L = 7.5D, 10D, 12.5D, and 15D downstream from the inlet face (\textit{refer to cross-sections P-P, Q-Q, R-R, and outlet} in Fig.\ref{Fig-sec}) for different inlet velocity conditions} 
\end{figure}

The flow profiles generated in the simulations demonstrate the ability of the constitutive model to capture the developing turbulence. The initial visible sign of instability originates in Case-A at a distance of $L=10D$ from the inlet face of the pipe (Fig. \ref{Fig-vx-06}). As the inlet velocity increases to 0.7 m/s and 0.9 m/s in Case-B and Case-C respectively, the local instability diffuses towards the inlet face resulting in turbulence much closer to the inlet (Figs. \ref{Fig-vx-07}-\ref{Fig-vx-09}). While in Fig. \ref{Fig-vx-06} the flow remains laminar up to $L=12.5D$ from the inlet, the laminar zone is predominantly restricted up to $L=7.5D$ in Fig. \ref{Fig-vx-07}. A further increase in the inlet velocity makes the flow turbulent even before $L=7.5D$ - Fig. \ref{Fig-vx-09}.

\section{\label{sec:5}Conclusion}
In this work, we presented results of numerical simulations of pipe flow based on Navier-Stokes theory with and without viscous strength. These results show that viscous strength is crucial for the appearance of material instabilities of the flow and is consistent with the experimental observation of Avila et al. \cite{avila2011onset}. Such material instabilities trigger local transitions to a chaotic flow that develops into turbulence. We hypothesize that material instabilities compete with the kinematic ones in the process of the transition to turbulence. Unfortunately, the classical Navier-Stokes model cannot capture the material instabilities and, thus, cannot describe the transition to turbulence in kinematically stable flows. Enhancement of the classical model with finite viscous strength opens new ways to simulate and understand the transition to turbulence.

\section*{Acknowledgements}
This work was supported by the Israel Science Foundation (ISF-394/20).

\section*{Author contributions statement}

K.V. conceived the idea,  S.K.L. conducted the numerical simulations, and S.K.L. and K.V. analysed the results.  All authors reviewed the manuscript.

\section*{Data Availability Statement}
The data sets used and analysed during the current study are available from the corresponding author on reasonable request.

\bibliographystyle{unsrtnat}
\bibliography{references}  






\end{document}